\documentclass[letterpaper]{ar-1col}

\usepackage[numbers]{natbib}
\usepackage{url}
\usepackage{verbatim}
\usepackage{amsmath, amssymb}
\usepackage{siunitx}
\usepackage{xspace}
\usepackage{xcolor}
\usepackage{ifoddpage}
\usepackage{epstopdf}

\newcommand{\eV}{{\rm eV}}

\newcommand{\lartpc}{LArTPC\xspace}
\newcommand{\lartpcs}{LArTPCs\xspace}
\newcommand{\uboone}{MicroBooNE\xspace}
\newcommand{\sbnd}{\text{SBND}\xspace}

\jname{Annu. Rev. Nucl. Part. Sci.}
\jvol{AA}
\jyear{2019}
\doi{10.1146/((please add article doi))}

\begin{document}

% Page header
\markboth{Machado $\bullet$ Palamara $\bullet$ Schmitz}{The Short-Baseline Neutrino Program}

% Title
\title{The Short-Baseline Neutrino Program at Fermilab}%

%Authors, affiliations address.
\author{Pedro A. N. Machado$^1$, Ornella Palamara$^1$ and \\ David W. Schmitz$^2$
\affil{$^1$Fermi National Accelerator Laboratory, Batavia, IL, USA, 60510}
\affil{$^2$Enrico Fermi Institute and Department of Physics, University of Chicago, Chicago, IL, USA, 60637}
}

%Abstract
\begin{abstract}
The Short-Baseline Neutrino, or SBN, program consists of three liquid argon time projection chamber detectors located along the Booster Neutrino Beam at the Fermi National Accelerator Laboratory. Its main goals include searches for new physics - particularly eV-scale sterile neutrinos, detailed studies of neutrino-nucleus interactions at the GeV energy scale, and the advancement of the liquid argon detector technology that will also be used in the DUNE/LBNF long-baseline neutrino experiment in the next decade. Here we review these science goals and the current experimental status of SBN. 
\end{abstract}

%Keywords, etc.
\begin{keywords}
SBN, neutrinos, short-baseline, sterile neutrinos, liquid argon time projection chamber, LArTPC
\end{keywords}
\maketitle

%Table of Contents
\tableofcontents

%%%%%%%%%%%%%%%%%%%%%%%%%%%%%%%%%%%%%%%%%%%%%%%%%%%%%%%%%%%%%%%%%%%%%%%%%%%%%%%%%%
\section{INTRODUCTION}
\label{sec:intro}
%%%%%%%%%%%%%%%%%%%%%%%%%%%%%%%%%%%%%%%%%%%%%%%%%%%%%%%%%%%%%%%%%%%%%%%%%%%%%%%%%%

The Short-Baseline Neutrino (SBN) program at Fermilab presents an exciting opportunity in experimental neutrino physics.  SBN will carry out precision searches for new physics in neutrinos and record millions of neutrino charged-current and neutral-current interactions on argon for untangling the physics of neutrino-nucleus scattering at the GeV energy scale.  In addition, SBN is providing a development platform for the liquid argon time projection chamber neutrino detector technology and has created a primary training ground for the international group of scientists and engineers working toward the flagship DUNE/LBNF long-baseline neutrino program in the US~\cite{Abi:2018dnh}.

SBN is designed to address the possible existence of \SI{1}{eV} mass-scale sterile neutrinos~\cite{Abazajian:2012ys}.  Sterile neutrino states, if they exist, are not directly observable since they do not interact with ordinary matter through the weak interaction, but active-sterile mixing could generate new oscillations among the standard neutrino flavors.
The search for light sterile neutrinos at SBN is motivated by a set of anomalous results in past neutrino data, most significantly from the LSND~\cite{Athanassopoulos:1996jb} and MiniBooNE~\cite{AguilarArevalo:2010wv,Aguilar-Arevalo:2013pmq,Aguilar-Arevalo:2018gpe} experiments.  We now require precision follow-up experiments to either confirm or rule out the existence of these new neutrino states.  SBN will test this important question using multiple, functionally identical detectors sitting along the same neutrino beam, which is the key to the experiment's world-leading sensitivity.  A discovery would reveal a new, unexpected form of fundamental particle and drive further experimentation in this area for years to come. A clear null result from SBN, or a Standard Model explanation for the earlier anomalies, would bring a welcome resolution to a long-standing puzzle and greatly clarify the current picture in neutrino physics. Light sterile neutrinos, in some scenarios, can have significant impacts on measurements of $CP$ violation and other oscillation parameters in DUNE (see e.g. Refs.~\cite{Gandhi:2015xza, Berryman:2015nua, Coloma:2017ptb, Choubey:2018kqq}) and on searches for neutrinoless double $\beta$ decay (see e.g. Refs.~\cite{Goswami:2005ng, Abada:2014nwa, Barea:2015zfa, Giunti:2015kza}), further motivating a precision search for sterile neutrinos at SBN.

\begin{figure}
 \checkoddpage
  \edef\side{\ifoddpage l\else r\fi}%
  \makebox[\textwidth][\side]{%
  \begin{minipage}{1.25\textwidth}
\centering
\includegraphics[width=\textwidth,trim={0mm 0mm 0mm 0mm},clip]{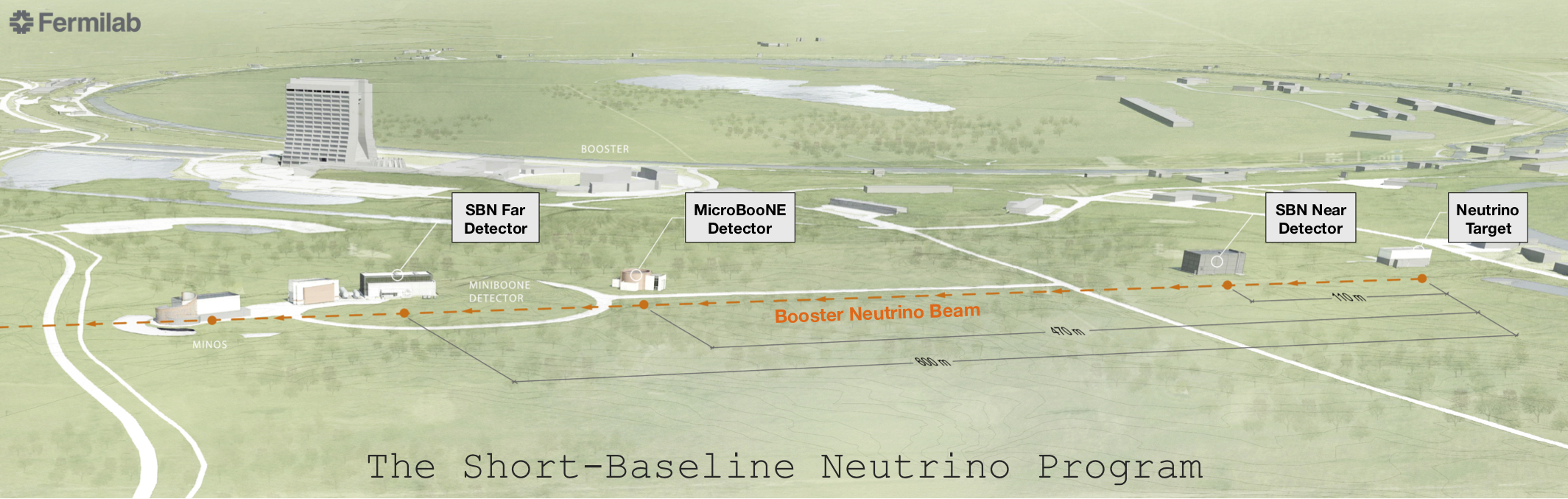}
\caption{An aerial view of the Short-Baseline Neutrino experimental area at Fermilab.  To the right is the neutrino beam target area where \SI{8}{GeV} protons from the Booster accelerator impinge a beryllium target.  The beam is focused along the orange dashed line (approximately \SI{7}{m} below grade) traveling toward the left (north).  The Near Detector, MicroBooNE, and Far Detector building locations are indicated. Image credit: Holabird \& Root.}
\label{fig:sbn}
\end{minipage}
}%
\end{figure}

Figure~\ref{fig:sbn} depicts the layout of the Short-Baseline Neutrino program, where three large liquid argon TPCs (\lartpc) will sit along the existing Booster Neutrino Beam (BNB) at the Fermi National Accelerator Laboratory in Illinois (USA).  The \uboone detector, an \SI{89}{ton} active mass \lartpc located \SI{470}{m} along the beam, has been collecting data in the BNB since October 2015. Earlier in that same year, a proposal~\cite{Antonello:2015lea} was presented and approved to augment the \uboone detector with two additional \lartpcs~-- a near detector close to the source that can characterize the neutrino beam before any substantial oscillation can occur and, thereby, greatly reduce systematic uncertainties in a search for oscillation signals downstream, and a larger far detector to be installed just downstream of \uboone to increase the statistics of a potential signal. The near detector, SBND (or the Short-Baseline Near Detector), will be an all new \SI{112}{ton} active mass \lartpc sited only \SI{110}{m} from the neutrino production target. The far detector will be the existing \SI{476}{ton} active mass ICARUS-T600 detector that has been refurbished and upgraded for optimal performance in SBN, and is now being readied for operation in a new experimental hall \SI{600}{m} from the target.  The addition of \sbnd and ICARUS creates a world-leading sterile neutrino search experiment that can cover the parameters allowed by past anomalies at $\geq 5\sigma$ significance.  ICARUS and SBND are scheduled to come online in 2019 and 2020, respectively.

In this Review, we summarize the current state of the sterile neutrino hypothesis in light of the latest experimental inputs and present the prospects for the SBN program to address this open question in the coming few years.  In Section~\ref{sec:steriles} we review the experimental anomalies (Sec.~\ref{sec:anomalies}), phenomenology (Sec.~\ref{sec:pheno}), and current experimental landscape (Sec.~\ref{sec:landscape}) of light sterile neutrinos. In Section~\ref{sec:sbn} we describe the beam and detectors of the Short-Baseline Neutrino experimental program, focusing on the new near and far detectors, SBND (Sec.~\ref{sec:sbnd}) and ICARUS (Sec.~\ref{sec:icarus}).  Then in Section~\ref{sec:science} we turn to the science program of SBN, beginning with the flagship sterile neutrino oscillation searches in Section~\ref{sec:osc}.  The science of SBN goes well beyond this, however, including a rich program of precision studies of the physics of neutrino-nucleus scattering (Sec.~\ref{sec:xsec}) and an ever-increasing range of ideas, now under active development, for using the SBN detectors and beam to search for signatures of new physics, from extra dimensions to light dark matter (Sec.~\ref{sec:bsm}).  We conclude in Section~\ref{sec:conclusion} with an outlook for the future at SBN.

%%%%%%%%%%%%%%%%%%%%%%%%%%%%%%%%%%%%%%%%%%%%%%%%%%%%%%%%%%%%%%%%%%%%%%%%%%%%%%%%%%
\section{LIGHT STERILE NEUTRINOS}
\label{sec:steriles}
%%%%%%%%%%%%%%%%%%%%%%%%%%%%%%%%%%%%%%%%%%%%%%%%%%%%%%%%%%%%%%%%%%%%%%%%%%%%%%%%%%

The sterile neutrino is a hypothetical particle, originally introduced by Bruno Pontecorvo in 1967 \cite{Pontecorvo:1967}, that does not experience any of the known fundamental forces, except gravity.  The sterile neutrino's existence must be indirectly observed, therefore, through its mixing with Standard Model neutrinos that could drive a new oscillation effect among the active flavors. Anomalies have existed in experimental data for more than 20 years that have been interpreted as hinting at just such an oscillation.  A definitive solution to this puzzle from a new generation of precision experiments is absolutely necessary, and this is one of the main goals of the SBN program.   

%%%%%%%%%%%%%%%%%%%%%%%%%%%%%%%%%%%%%%%%%%%%%%%%%%%%%%%%%%%%
\subsection{The Experimental Anomalies}
\label{sec:anomalies}
%%%%%%%%%%%%%%%%%%%%%%%%%%%%%%%%%%%%%%%%%%%%%%%%%%%%%%%%%%%%

By now, numerous neutrino experiments have established the existence of flavor oscillations between the three neutrinos predicted by the Standard Model (see, e.g. Ref.~\cite{GonzalezGarcia:2007ib}). These oscillations are characterized by the presence of the so-called solar and atmospheric mass splittings, $\Delta m^2_{\odot}\simeq7.4\times 10^{-5}~\eV^2$ and $|\Delta m^2_{\rm atm}|\simeq 2.5\times10^{-3}~\eV^2$, as well as three mixing angles, $\theta_{12}=34^\circ$, $\theta_{13}=8.6^\circ$, and $\theta_{23}\sim 45^\circ$~\cite{Esteban:2018azc}. Still, for the last two decades, several experimental results seem to  indicate the existence of an additional neutrino without electroweak interactions -- \emph{a sterile neutrino}. 

The Liquid Scintillator Neutrino Detector (LSND) at the Los Alamos National Laboratory consisted of a stopped pion source producing an intense flux of $\bar\nu_\mu$ with energies up to \SI{53}{MeV}. A liquid scintillator detector, located roughly \SI{30}{m} from the source, was optimized to observe electron neutrino events via the inverse beta decay process in carbon, $\bar\nu_e\,p\to e^+n$, by detecting the Cherenkov and scintillation light produced by the $e^+$ and the delayed \SI{2.2}{MeV} photon from  neutron capture. The main backgrounds at LSND were conventional $\bar\nu_e$ production in the beam stop and $\pi^-$ decay in flight followed by $\bar\nu_\mu \,p\to\mu^+n$ where the $\mu^+$ is mis-identified as an $e^+$.  LSND has observed an excess of $87.9\pm22.4\pm6.0$ $\bar\nu_e$ events over these backgrounds, a $3.8\sigma$ deviation from expectations~\cite{Aguilar:2001ty}. Interpreting the LSND result in a neutrino oscillation formalism leads to an additional mass-squared splitting $\Delta m^2 \geq \mathcal{O}(\SI{0.1}{eV^2})$, thus requiring physics beyond the Standard Model. This is the origin of what is referred to as the short-baseline neutrino anomaly.

Following LSND, the MiniBooNE experiment was proposed to test the sterile neutrino hypothesis, and it is still running today. 
MiniBooNE is located on the Boster Neutrino Beam at Fermilab, which peaks at $\sim$\SI{700}{MeV} neutrino energy and has a magnetic horn system allowing for focusing negatively or positively charged mesons leading to a mostly $\nu_\mu$ or $\bar\nu_\mu$ neutrino beam. A mineral oil detector optimized to observe Cherenkov light emitted by electrons and muons is located \SI{540}{m} downstream from the neutrino production target. The different energy configuration and event signature makes MiniBooNE backgrounds very different from those in LSND.  However, the higher energy and longer baseline make it sensitive to the same range of $L/E$, ensuring that MiniBooNE probes a mass squared splitting of $\mathcal{O}(1~\eV^2)$, similarly to LSND. 

Since the event classification relies on the Cherenkov ring topology, electrons and photons are indistinguishable in MiniBooNE. The main backgrounds are the following: (1)~$\pi^0$ mis-identification as an electron-like event; (2)~$\nu_e$ from kaon and muon decays in the beamline; (3)~single photon production via the resonant process $\Delta\to N\gamma$; and (4)~single photon events from neutrino interactions in the dirt and material surrounding the detector. Background (3) has been the focus of particular interest as it does not come directly from measurements, but rather is based on non-perturbative theoretical calculations, see e.g. Refs.~\cite{Hill:2010zy, Wang:2014nat}. 

After collecting $12.84~(11.27)\times 10^{20}$ protons on target in neutrino (anti-neutrino) modes, the MiniBooNE collaboration has observed excesses of electron-like events in both modes, leading to a $4.7\sigma$ deviation from the expected background~\cite{Aguilar-Arevalo:2013pmq, Aguilar-Arevalo:2018gpe}. If interpreted as neutrino oscillations, the excesses of events and their energy distributions in  MiniBooNE  and LSND are compatible, strengthening the short-baseline anomaly. 

Other short-baseline neutrino anomalies have been reported in the $\nu_e$ and $\bar\nu_e$ disappearance modes, both in the detection of neutrinos from nuclear power reactors and in calibration runs of solar neutrino experiments using radioactive sources. 
A 2011 reevaluation of the flux of neutrinos produced in reactors lifted the expected $\bar\nu_e$ flux by $\sim3\%$~\cite{Mention:2011rk, Huber:2011wv}. This effect, together with improved theoretical uncertainties, led to a shift in the ratio of total observed events over the predicted number of events in a number of reactor experiments, with an average value of $R=0.94\pm0.02$. This is the origin of the so-called `reactor anomaly'. Some have noted that a possible underestimation of theoretical uncertainties could have considerably increased the significance of the reactor anomaly~\cite{Hayes:2013wra}, although the NEOS~\cite{Ko:2016owz} and DANSS~\cite{Alekseev:2018efk} experiments have observed spectral features also consistent with sterile neutrino oscillations.  Additionally, calibration data from gallium solar neutrino detectors using intense radioactive neutrino sources~\cite{Hampel:1998xg, Abdurashitov:2005ax} and the theoretical cross section for neutrino capture {$\nu_e+^{71}$Ga$\,\to^{71}$Ge$+e^-$}~\cite{Bahcall:1997eg} leads to a $3\sigma$ deficit compared to the expected number of events~\cite{Acero:2007su, Giunti:2010zu, Kopp:2013vaa}. This is known as the `gallium anomaly'.

In view of these unexpected results, it is vital to characterize and evaluate the feasibility of sterile neutrinos as an explanation of the short-baseline experimental anomalies. In the next sections we present the general phenomenology of light sterile neutrinos and then summarize the current experimental landscape.

%%%%%%%%%%%%%%%%%%%%%%%%%%%%%%%%%%%%%%%%%%%%%%%%%%%%%%%%%
\begin{textbox}[h]\section{Summary of the short-baseline neutrino anomalies}
%%%%%%%%%%%%%%%%%%%%%%%%%%%%%%%%%%%%%%%%%%%%%%%%%%%%%%%%%

Four main `anomalies' have been observed in neutrino experiments at short-baseline, consistent with the mixing of the standard neutrinos with a fourth, non-weakly-interacting `sterile' species -- the data could be indicating a heavier, mostly-sterile mass state with mass splittings $\Delta m_{43}^2 \approx \Delta m_{42}^2 \approx \Delta m_{41}^2\sim\mathcal{O}(\SI{1}{eV^2})$.

\vspace{0.5em}
\noindent {\bf $\bullet$~LSND:} Stopped pion source with a detector optimized to probe $\bar\nu_e$ via inverse beta decay. A $3.8\sigma$ excess of events over backgrounds was observed, compatible with $\bar\nu_{\mu} \to \bar\nu_e$ oscillations with $L/E \approx \SI{1}{m/MeV}$~\cite{Aguilar:2001ty}.

\vspace{0.5em}
\noindent {\bf $\bullet$~MiniBooNE:} Accelerator neutrino source with the capability of producing a dominant $\nu_\mu$ or $\bar\nu_\mu$ beam.  Excesses of $\nu_e$($\bar\nu_e$) events in $\nu_\mu$($\bar\nu_\mu$) mode were observed over backgrounds, amounting to a $4.5\sigma(2.8\sigma)$ discrepancy from expectations. The observed excesses are found to be compatible with LSND within a sterile neutrino framework~\cite{Aguilar-Arevalo:2018gpe}.

\vspace{0.5em}
\noindent {\bf $\bullet$~The `Reactor anomaly':} A reevaluation of the $\bar\nu_e$ fluxes from nuclear reactors with improved theoretical uncertainties led to a deficit in many past experiments in the total number of events with respect to theoretical expectations at the $3\sigma$ level~\cite{Mention:2011rk, Huber:2011wv}. The size of these theoretical uncertainties has been under debate~\cite{Hayes:2013wra}. More recently, some spectral features have been observed consistent with sterile neutrino oscillations with $\Delta m^2\sim$\,eV$^2$~\cite{Ko:2016owz, Alekseev:2018efk}.

\vspace{0.5em}
\noindent {\bf $\bullet$~The `Gallium anomaly':} An overall deficit in the number of $\nu_e$ events from radioactive sources~\cite{Hampel:1998xg, Abdurashitov:2005ax} with respect to theoretical expectations~\cite{Bahcall:1997eg} at the $3\sigma$ level was seen during calibration runs of solar neutrino experiments~\cite{Acero:2007su, Giunti:2010zu, Kopp:2013vaa}.
\end{textbox}

%%%%%%%%%%%%%%%%%%%%%%%%%%%%%%%%%%%%%%%%%%%%%%%%%%%%%%%%%%
\subsection{Phenomenology of Sterile Neutrinos}
\label{sec:pheno}
%%%%%%%%%%%%%%%%%%%%%%%%%%%%%%%%%%%%%%%%%%%%%%%%%%%%%%%%%%

To better understand the status of eV-scale sterile neutrino physics, we introduce the formalism and notation used hereafter. We will focus on the scenario where one sterile neutrino is added to the neutrino spectrum, the 3+1 scenario, and we will comment on scenarios with additional sterile neutrinos later.

Neutrino oscillations require non-zero and non-degenerate neutrino masses, as well as the presence of mixing. Neutrino mixing amounts to the fact that the eigenstates produced by electroweak interactions (flavor states) are non-trivial linear combinations of mass eigenstates, that is, eigenstates of the free Hamiltonian with well defined masses. A 3+1 mixing matrix:
\begin{equation}
    \begin{pmatrix}
    \nu_e \\ 
    \nu_\mu \\
    \nu_\tau \\
    \nu_s \\
    \end{pmatrix}
    =
    \begin{pmatrix}
    U_{e1} & U_{e2} & U_{e3} & U_{e4} \\ 
    U_{\mu1} & U_{\mu2} & U_{\mu3} & U_{\mu4} \\
    U_{\tau1} & U_{\tau2} & U_{\tau3} & U_{\tau4} \\
    U_{s1} & U_{s2} & U_{s3} & U_{s4} \\
    \end{pmatrix}
    \begin{pmatrix}
    \nu_1 \\ 
    \nu_2 \\
    \nu_3 \\
    \nu_4 \\
    \end{pmatrix}    
\end{equation}
can be parameterized as (see, e.g., Ref.~\cite{Dentler:2018sju})
\begin{equation}
  U = R_{34}(\theta_{34})
	  R_{24}(\theta_{24},\delta_{24})
	  R_{14}(\theta_{14})
	  R_{23}(\theta_{23})
	  R_{13}(\theta_{13},\delta_{13})
	  R_{12}(\theta_{12},\delta_{12}),
\end{equation}
 where $R_{ij}$ denotes a rotation in the $ij$-plane by an angle $\theta_{ij}$ and a possible phase $\delta_{ij}$ (if present). The standard three neutrino framework can be recovered by setting $\theta_{i4}=0$ for $i=1,2,3$ 
 and identifying $\delta_{13}$ with the 3 neutrino phase typically denoted by $\delta_{CP}$ (in this case, $\delta_{12}$ becomes unphysical).

As long as $\Delta m^2_{41}\gg |\Delta m^2_{31}|,\Delta m^2_{21}$, oscillations at short-baseline experiments can be well described by a two-flavor vacuum oscillation formula
\begin{equation}
  P_{\alpha \beta} = \delta_{\alpha\beta} - 4|U_{\alpha\beta}|^2(\delta_{\alpha\beta}-|U_{\alpha\beta}|^2)\sin^2\left(\frac{\Delta m^2_{41} L}{4E}\right),
\end{equation}
where $L$ is the baseline and $E$ is the neutrino energy.

In a nutshell, each  oscillation channel $\nu_\alpha\to\nu_\beta$ is driven by a different effective mixing $\theta_{\alpha \beta}$, namely,
\begin{align}
  \nu_\mu\to\nu_e: &\,\, \sin^22\theta_{\mu e}\equiv 4|U_{\mu4}|^2|U_{e4}|^2 &  \text{(LSND, MiniBooNE anomalies);} \label{eq:APP}\\
  \nu_e\to\nu_e: & \,\,\sin^22\theta_{e e}\equiv 4|U_{e4}|^2(1-|U_{e4}|^2) & \text{(Reactor, Gallium anomalies);}\\
  \nu_\mu\to\nu_\mu: & \,\,\sin^22\theta_{\mu\mu}\equiv 4|U_{\mu 4}|^2(1-|U_{\mu 4}|^2) & \text{(no anomaly observed).}\label{eq:DIS-mu}
\end{align}
There are two important features that should be noticed regarding sterile neutrino oscillations. First, the characteristic $\sin^2(\Delta m^2 L/4E)$ dependence of neutrino oscillations may allow to distinguish it from other possible explanations of the anomalies.
Second, short-baseline transitions $\nu_\mu\to\nu_e$  require non-zero $U_{e4}$ and $U_{\mu4}$, thus necessarily inducing both $\nu_\mu\to\nu_\mu$ and $\nu_e\to\nu_e$ probabilities less than 1. This can be used to over-constrain the parameter space by observing $\nu_e$ appearance together with $\nu_e$ and $\nu_\mu$ disappearance at short-baselines.
As we will see shortly, these two features will be crucial to test the sterile neutrino hypothesis.

%%%%%%%%%%%%%%%%%%%%%%%%%%%%%%%%%%%%%%%%%%%%%%%%%%%%%%%%%%%%%
\subsection{Light Sterile Neutrino Experimental Landscape in 2019}
\label{sec:landscape}
%%%%%%%%%%%%%%%%%%%%%%%%%%%%%%%%%%%%%%%%%%%%%%%%%%%%%%%%%%%%%

The current experimental status of eV-scale sterile neutrinos is puzzling. While experimental data supporting the existence of eV-scale sterile neutrinos with non-negligible mixing with the active species continue to be amassed, several experiments that should be sensitive to such mixing have not observed anything beyond the three neutrino oscillation framework. 

Neutrino experimental data relevant to the short-baseline anomaly can naturally be divided into three groups: $\nu_e$ appearance, $\nu_e$ disappearance, and $\nu_\mu$ disappearance. Instead of providing an extensive overview of all experimental results, we will focus here on the most sensitive experiments in each data set and the complementarity between these data sets. For more details, we refer the reader to Refs.~\cite{Dentler:2018sju, Collin:2016rao, Gariazzo:2017fdh}. All numbers quoted below for the limits on preferred regions refer to 99\% C.L.

We start with tests of $\nu_e$ appearance, which is driven by $\sin^2 2\theta_{\mu e}$, see Eq.~(\ref{eq:APP}). The OPERA~\cite{Agafonova:2013xsk} and ICARUS~\cite{Antonello:2012pq} experiments running in the CERN to Gran Sasso neutrino beam constrain $\sin^22\theta_{\mu e}<0.015$, roughly independent of $\Delta m^2$. Taking these into account, the two anomalous results coming from MiniBooNE and LSND require $2\times 10^{-3}\lesssim\sin^22\theta_{\mu e}\lesssim0.015$ and $\Delta m^2$ above \SI{0.3}{eV^2}.  In addition, values of $\Delta m^2$ larger than about \SI{1.5}{eV^2} with large enough mixing angle to explain the anomalies are disfavored by the KARMEN~\cite{Armbruster:2002mp} and NOMAD~\cite{Astier:2003gs} experiments.

The other channel that presents anomalous signals is $\nu_e$ disappearance. As discussed above, many short-baseline reactor experiments observe a deficit of events with respect to theoretical expectations following an updated estimate of reactor fluxes. Two recent results stand out; NEOS~\cite{Ko:2016owz} and DANSS~\cite{Alekseev:2018efk} each see mild indications of oscillations in the energy spectrum and, for the latter, in multiple baseline configurations. Null results from solar neutrino data and the LSND/KARMEN $^{12}$C data constrain $|U_{e4}|^2\lesssim 0.065$. In global data fits, short-baseline reactor data and the Gallium calibration experiments indicate a $3.2\sigma$ preference for non-zero values of $|U_{e4}|^2$, with a best fit at $\Delta m^2 = \SI{1.3}{eV^2}$ and $|U_{e4}|^2=0.009$. These numbers are obtained taking a free normalization for the antineutrino flux from each of the isotopes in the reactors. This continues to be a very active area of experimentation~($\nu_e$/$\bar\nu_e$~dis.), and one that is highly complementary to the searches that will take place at SBN~($\nu_e$ app.~and $\nu_\mu$ dis.).  The STEREO~\cite{Almazan:2018wln} and PROSPECT~\cite{Ashenfelter:2018iov} experiments are ongoing, and the SoLid~\cite{Abreu:2017bpe} experiment will soon be taking data. These new short-baseline reactor experiments will significantly contribute to a resolution of the reactor anomaly. 

Finally, no deviation from the three neutrino framework has been observed in the $\nu_\mu$ disappearance data sets. Here, the constraints on $|U_{\mu4}|^2$ are driven by MINOS/MINOS+~\cite{Adamson:2017uda}, the IceCube atmospheric neutrino sample~\cite{TheIceCube:2016oqi}, CDHS~\cite{Dydak:1983zq}, and the MiniBooNE/SciBooNE disappearance analyses~\cite{AguilarArevalo:2009yj, Cheng:2012yy}. MINOS/MINOS+ performs a fit to their near~($\sim$\SI{1}{km}) and far~($\sim$\SI{735}{km}) detector data in a broad band beam with neutrino energies from 1-\SI{40}{GeV}, thus constraining a large region in $\Delta m^2$. The IceCube constraint, on the other hand, relies on the non-observation of the presence of a matter driven resonance in the atmospheric anti-neutrino spectrum at TeV energies. Both CDHS and MiniBooNE perform short-baseline oscillation searches. The combined limit derived from these experiments is $|U_{\mu4}|^2\lesssim 0.008$ for $0.1\lesssim\Delta m^2\lesssim10$~eV$^2$.

\begin{figure}
\centering
\includegraphics[width=0.71\textwidth,trim=0mm 0mm 0mm 20mm]{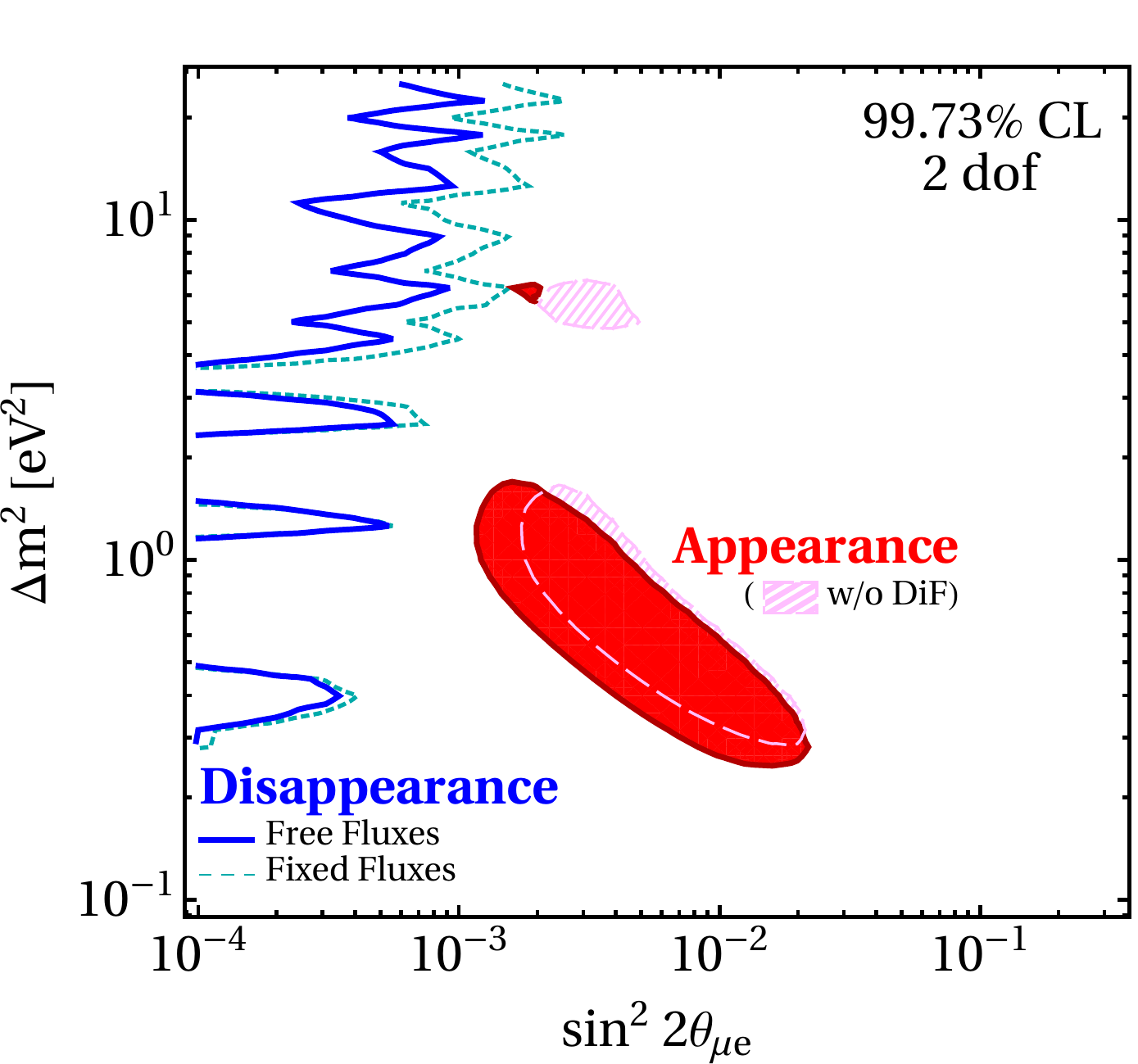}
\caption{Preferred regions in the $\sin^22\theta_{\mu e}\equiv 4|U_{e4}|^2|U_{\mu4}|^2$ versus $\Delta m^2$ plane for the disappearance data set using free reactor fluxes (solid blue) or fixed reactor fluxes (dashed blue), as well as for the appearance data set with (red) and without (hatched red) the LSND decay in flight sample. The curves are drawn at the 99.73\% C.L. for 2 degrees of freedom. Figure taken from Ref.~\cite{Dentler:2018sju}.} \label{fig:tension}
\end{figure}

With these results at hand, a global analysis of the sterile neutrino hypothesis can be made. Taking the $\nu_\mu$ disappearance data and the mild constraint from solar neutrinos and the LSND/KARMEN $^{12}$C data, we see that $\sin^2 2\theta_{\mu e}\equiv4|U_{e4}|^2|U_{\mu4}|^2 \lesssim 0.002$, already uncovering a tension between these data sets and the MiniBooNE/LSND anomaly. A more sophisticated analysis can be performed using a parameter goodness of fit test to compare all  disappearance data against the appearance results~\cite{Dentler:2018sju}. The allowed regions for these two data sets have no overlap at the $3\sigma$ level, as can be seen in Fig.~\ref{fig:tension}, representing a $4.7\sigma$ tension\footnote{While this result has been derived using the 2012 MiniBooNE data release, the recent results~\cite{Aguilar-Arevalo:2018gpe} are expected to slightly worsen the tension.}. This tension is quite robust in the sense that it does not depend strongly on the overall reactor neutrino flux normalization or on the constraints derived from any single experiment.

A few additional points are worth noting. While this global analysis has been performed for a 3+1 scenario, no significant improvement is expected in scenarios with extra sterile neutrino species (see, e.g., Refs.~\cite{Kopp:2013vaa, Conrad:2012qt}). Also, light sterile neutrinos with sizable mixing angles (above the percent level, see e.g. Ref~\cite{Archidiacono:2013xxa}) would thermalize in the early universe, contributing to the effective number of relativistic species. The current limit, $N_{\rm eff}=2.99\pm0.17$, derived from observations of the cosmic microwave background and fits to the $\Lambda$CDM model of cosmology~\cite{Aghanim:2018eyx}, would imply a non-standard cosmological history for the sterile neutrinos and thus more ingredients of beyond Standard Model physics. Finally, as we will see shortly, SBN will play a major role in revealing the origin of the short-baseline anomalies. On one hand, SBN will be very sensitive to both $\nu_\mu\to\nu_e$ appearance and $\nu_\mu\to\nu_\mu$ disappearance channels, directly testing  the LSND/MiniBooNE anomaly and having an important impact on the tension between data sets. On the other hand, the $e$-$\gamma$ discrimination capability of LArTPCs will be important to understand the signal/background nature of the $\nu_e$-like event excess observed by MiniBooNE.

%%%%%%%%%%%%%%%%%%%%%%%%%%%%%%%%%%%%%%%%%%%%%%%%%%%%%%%%%%%%%%%%%%%%%%%%%%%%%%%%%
\section{THE SBN EXPERIMENTAL PROGRAM}
\label{sec:sbn}
%%%%%%%%%%%%%%%%%%%%%%%%%%%%%%%%%%%%%%%%%%%%%%%%%%%%%%%%%%%%%%%%%%%%%%%%%%%%%%%%%

Key to the science goals of SBN is the use of multiple detectors that will measure the same neutrino beam at different distances from the source.  Further, while not identical designs, the use of functionally identical liquid argon TPCs (see text box on the next page) is critical to minimizing systematic uncertainties when comparing event distributions at the different locations along the beam and searching for oscillation signals. Below is briefly introduced the neutrino source for SBN and the new \lartpcs currently being prepared for the SBN program, SBND and ICARUS.  The \uboone detector has been operating since 2015 and is described in detail in Ref.~\cite{Acciarri:2016smi}.

%%%%%%%%%%%%%%%%%%%%%%%%%%%%%%%%%%%%%%%%%%%%%%%%%%%%%%%%%%%%%%
\subsection{The Booster Neutrino Beam at Fermilab}
\label{sec:bnb}
%%%%%%%%%%%%%%%%%%%%%%%%%%%%%%%%%%%%%%%%%%%%%%%%%%%%%%%%%%%%%%

The SBN program makes use of the Booster Neutrino Beam (BNB) at Fermilab. The neutrino beam is created by extracting \SI{8}{GeV} kinetic energy protons from the Booster accelerator and impacting them on a beryllium target to produce a secondary beam of hadrons, mainly pions. Charged secondaries are focused by a single toroidal aluminum alloy focusing horn that surrounds the target. The horn is supplied with \SI{174}{kA} in \SI{143}{\mu s} pulses coincident with proton delivery. The horn can be pulsed with either polarity, thus focusing either positives or negatives and de-focusing the other. Focused mesons are allowed to propagate down a \SI{50}{m} long, \SI{0.91}{m} radius air-filled tunnel where the majority will decay to produce muon and electron neutrinos. The remainder are absorbed into a concrete and steel absorber at the end of the \SI{50}{m} decay region. The Booster spill length is \SI{1.6}{\mu s} with nominally 5$\times$10$^{12}$ protons per spill delivered to the beryllium target.  The BNB has already successfully and stably operated for more than fifteen years in both neutrino and anti-neutrino modes.  The fluxes are well understood thanks to a detailed simulation~\cite{AguilarArevalo:2008yp} developed by the MiniBooNE collaboration and the availability of dedicated hadron production data for \SI{8.9}{GeV/c} p+Be interactions collected at the HARP experiment at CERN~\cite{Catanesi:2007ab}. Systematic uncertainties associated with the beam have also been characterized in a detailed way, with a total error of ∼9\% at the peak of the $\nu_\mu$ flux and larger in the low and high energy regions. The composition of the flux in neutrino mode (focusing positive hadrons) is energy dependent, but is dominated by $\nu_\mu$~(93.6\%), followed by $\bar{\nu}_\mu$~(5.9\%), with an intrinsic $\nu_e/{\bar{\nu_e}}$ contamination at the level of 0.5\% at energies below \SI{1.5}{GeV}.

\begin{figure}[h]
\centering
\includegraphics[width=1.13\textwidth,trim=0mm 0mm 0mm 0mm,clip]{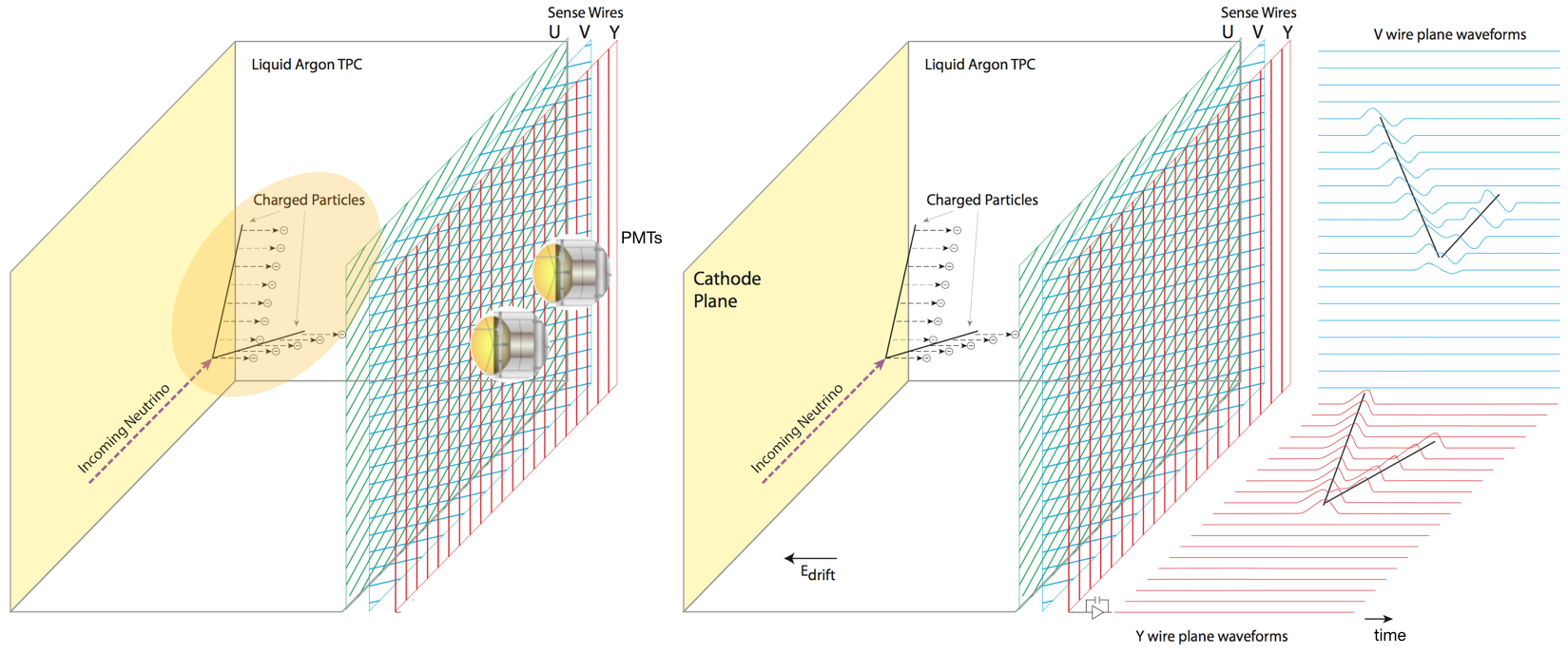}
\caption{Operating principle of the liquid argon time projection chamber neutrino detector. 
}
\label{fig:tpc}
\end{figure}

\begin{textbox}[t]
\section{The liquid argon time projection chamber neutrino detector}

The concept of the \lartpc for neutrino detection was first proposed by Carlo Rubbia in 1977~\cite{Rubbia:1977zz}. The working principle of the device is illustrated in Fig.~\ref{fig:tpc}. The detector consists of a large open volume of ultra-pure liquid argon (\SI{87}{K}) surrounded by a high voltage cathode surface on one face and an anode surface opposite to it. When a neutrino undergoes a charged-current or neutral-current weak interaction with an argon nucleus, resulting charged particles ionize and excite argon atoms as they propagate in the liquid.  Freed electrons drift in the inert medium under the influence of the electric field between the cathode and anode (made uniform by a field cage surrounding the liquid volume), with field strengths of ∼\SI{500}{V/cm} typical. In a `single-phase' wire readout detector, the clouds of drifted electrons generate small currents on taught sense wires located on the anode side of the detector boundary.  The wires are closely spaced to form planes (wire pitch in the 3-\SI{5}{mm} range). To generate multi-dimensional views of particle tracks, two or three planes of wires, oriented at different angles are used. The wires are biased to guide electrons past the first wire planes, where the electron clouds induce bipolar signals as they pass, and they collect on the positively biased back most plane, producing a unipolar signal proportional to the total ionization in that location. The detector, therefore, is a totally active, fine-sampling calorimeter with millimeter-level particle tracking capabilities.  Electron drift speeds are slow, in the range of \SI{1.6}{mm/\mu s}, requiring a continuous readout time of 1-\SI{2}{milliseconds} for a detector that is 2-\SI{3}{m} across.

The ionized and excited argon atoms also form short-lived argon dimers that decay and emit scintillation light in the vacuum ultraviolet ($\lambda = \SI{128}{nm}$, E = \SI{9.69}{eV}).  To detect this light with conventional photosensors, light collection systems in LAr typically rely on a wavelength shifting compound such as tetraphenyl butadiene (TPB) to down-convert the VUV scintillation photons into the visible.  The light signal is critical to the operation of the TPC as it provides the $t_0$ for an interaction within the volume and, therefore, a measure of the position along the drift direction.  In detectors with sufficient light collection efficiency, the photon signal can also contribute to 3D reconstruction and calorimetry measurements.  
\end{textbox}

%%%%%%%%%%%%%%%%%%%%%%%%%%%%%%%%%%%%%%%%%%%%%%%%%%%%%%%%%%
\subsection{The SBN Near Detector: SBND}
\label{sec:sbnd}
%%%%%%%%%%%%%%%%%%%%%%%%%%%%%%%%%%%%%%%%%%%%%%%%%%%%%%%%%%

\begin{figure}
    \begin{minipage}{0.38\textwidth}
        \centering
        \includegraphics[width=1\textwidth,trim=0mm 0mm 0mm 10mm,clip]{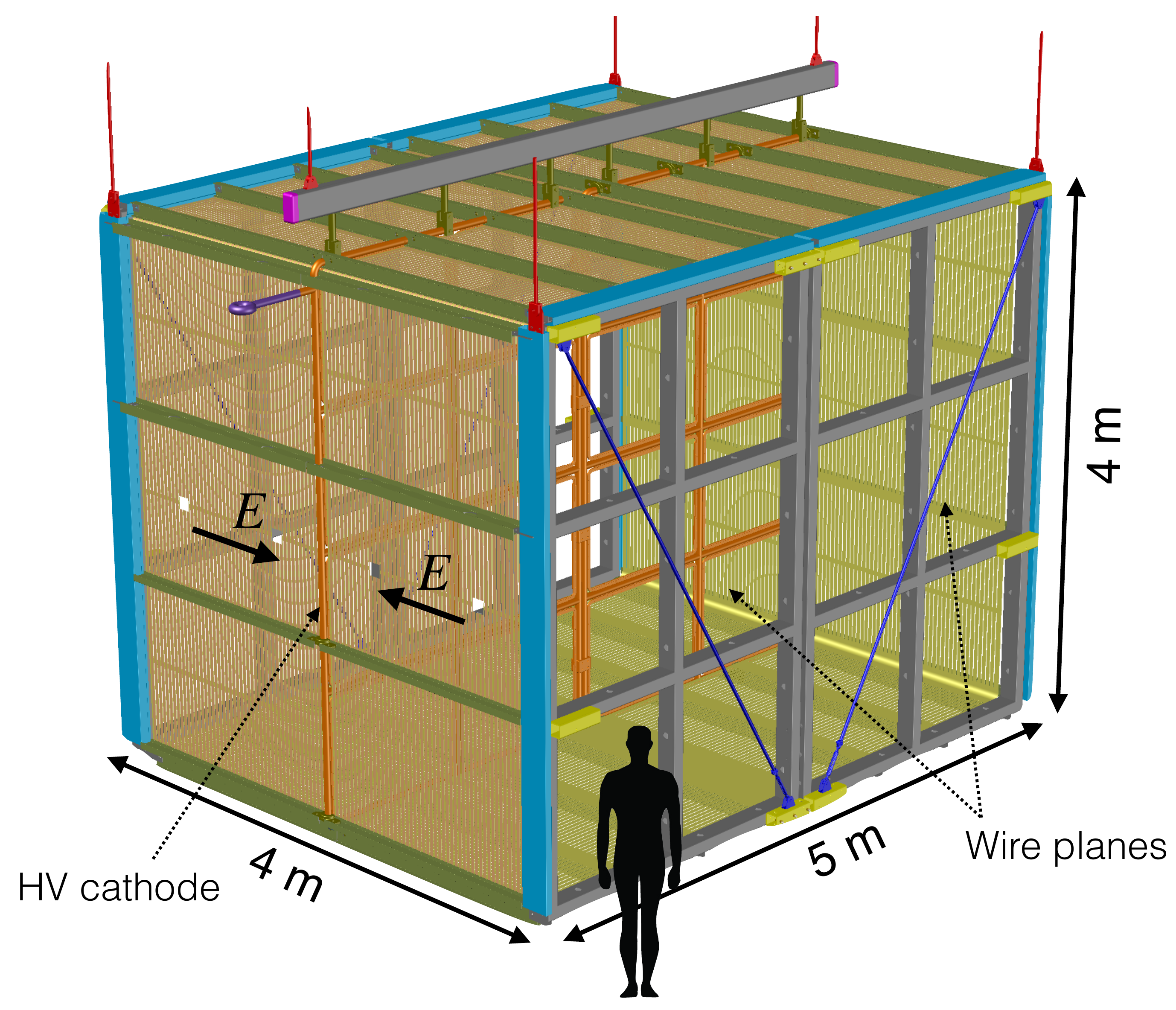}
        \includegraphics[width=0.85\textwidth,trim=0mm 0mm 0mm 0mm,clip]{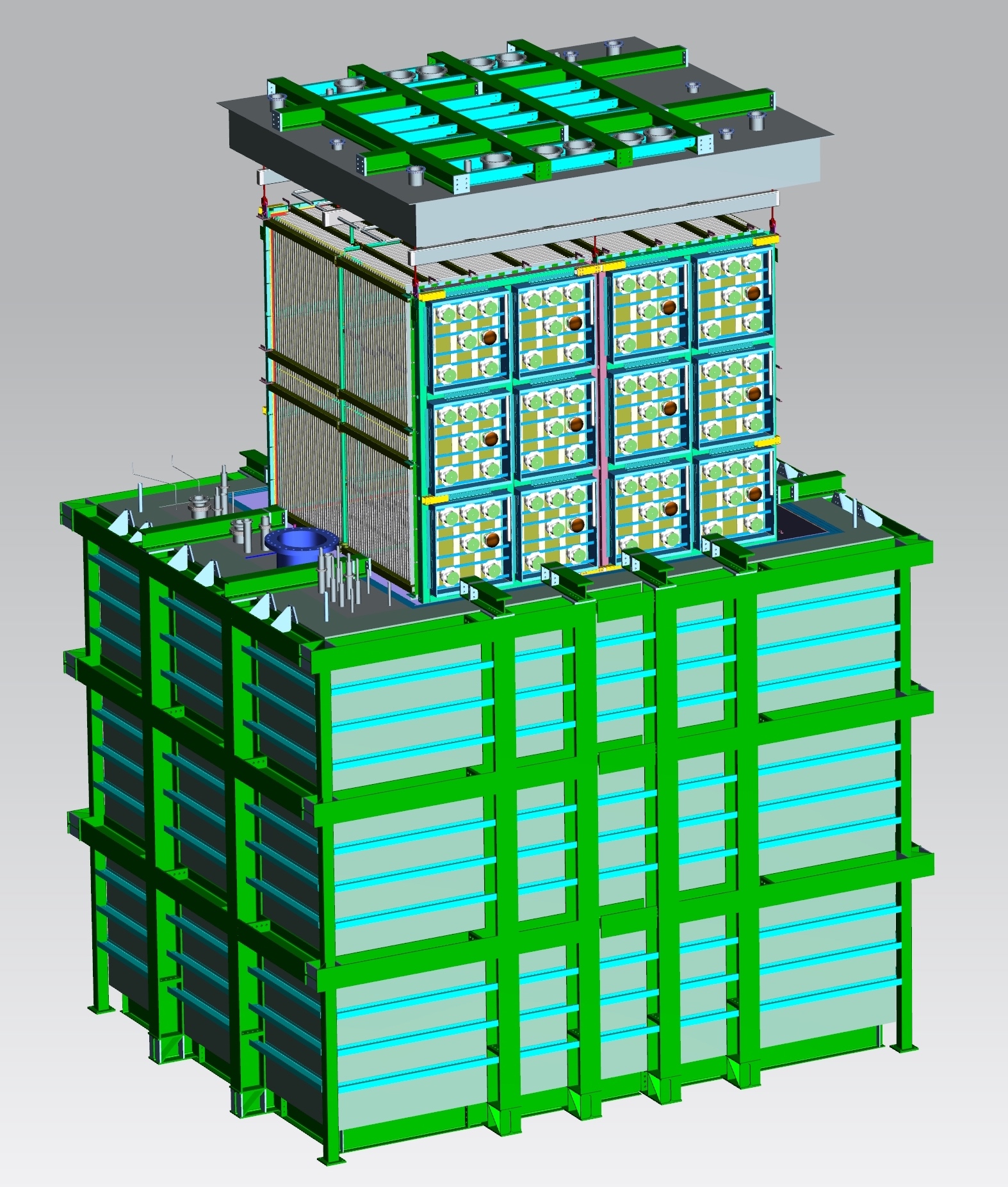}
    \end{minipage}
    \qquad 
    \begin{minipage}{0.515\textwidth}
        \centering
        \includegraphics[width=1\textwidth,trim=0mm 0mm 0mm 0mm,clip]{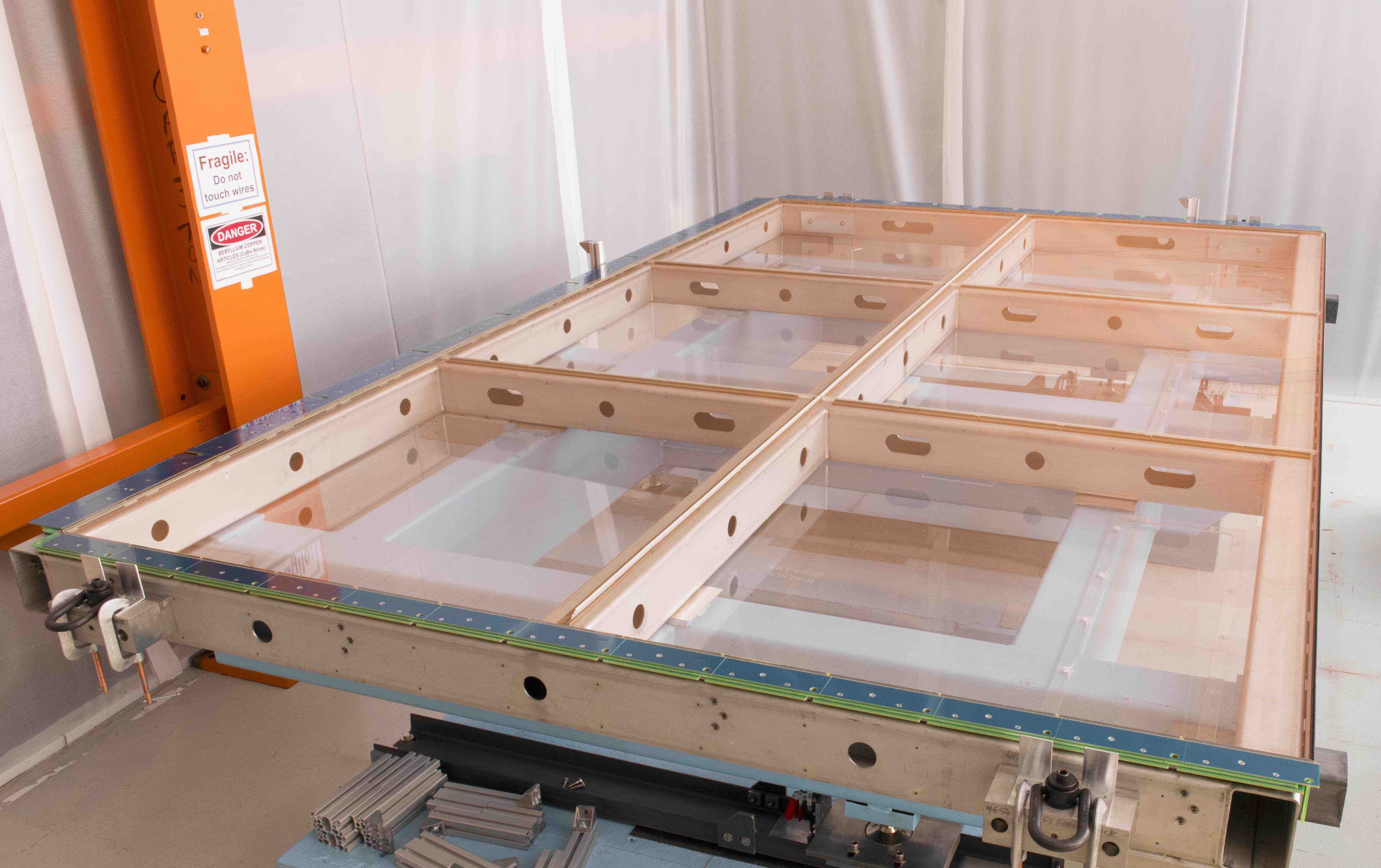}\\
        \vspace{-2mm}
        \includegraphics[width=1\textwidth,trim=0mm 0mm 0mm 0mm,clip]{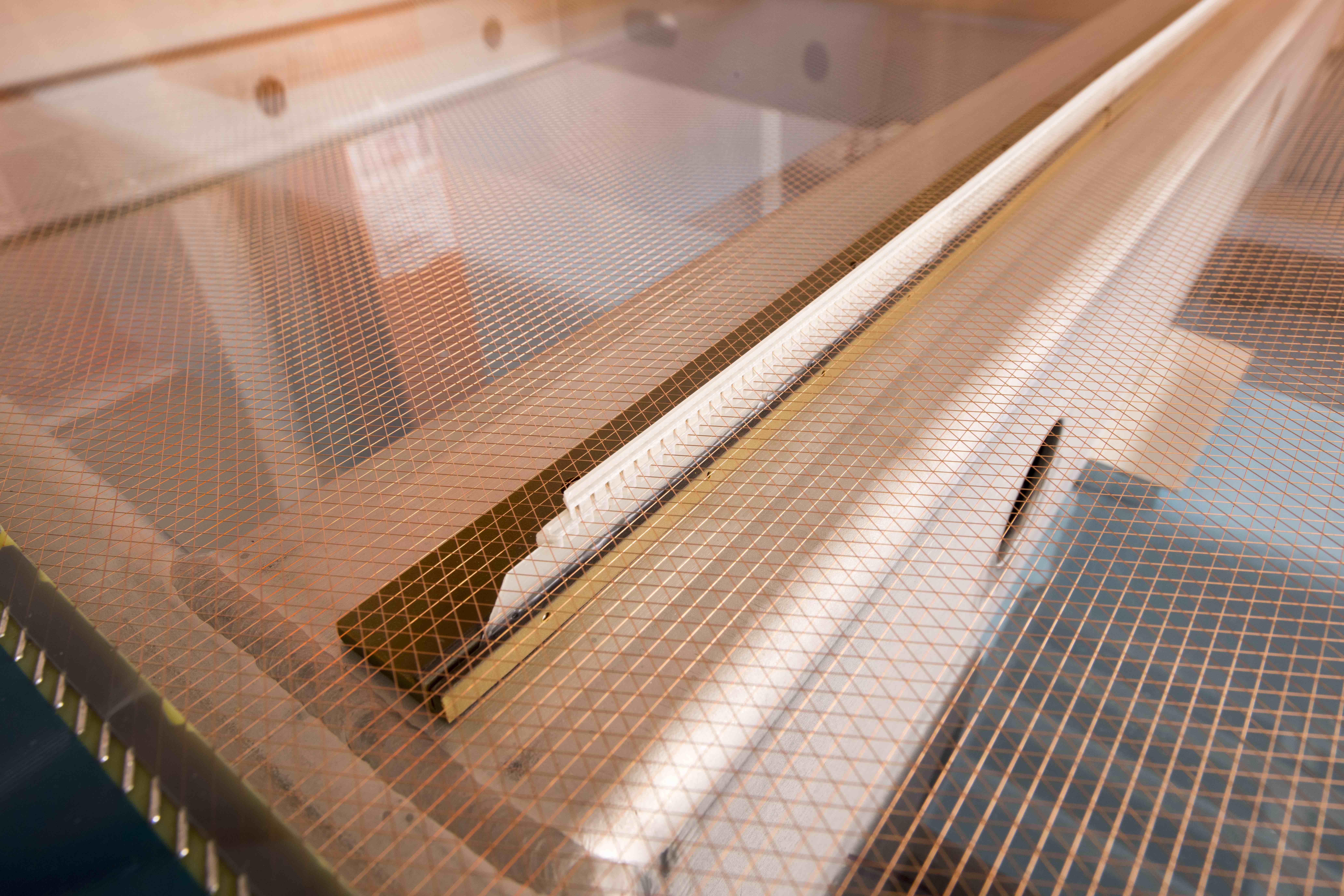}
    \end{minipage}
    \caption{{\it Left:}~The SBN near detector, SBND, contains two TPC drift regions on either side of a central high voltage cathode.  Each \SI{2}{m} drift region is read out by a pair of coupled anode plane assemblies, each approximately \SI{4}{m} tall by \SI{2.5}{m} wide. The TPC and photon detectors are suspended from a removable cryostat lid. The TPC is shown just before being lowered into place inside the membrane cryostat. {\it Right:}~One of four completed SBND anode planes. Each assembly contains more than 2800 copper-beryllium readout wires. Image credit:~Reidar Hahn, Fermilab.}
    \label{fig:sbnd}
\end{figure}

The Short-Baseline Near Detector (\sbnd), a \SI{112}{ton} LArTPC, is the near detector in the SBN program and will measure the unoscillated BNB neutrino flux. \sbnd is under construction and will be placed in a new experimental hall located \SI{110}{m} downstream from the BNB target. \sbnd has an active volume of $\SI{5.0}{m}~(\mathrm{L})\times\SI{4.0}{m}~(\mathrm{W})\times\SI{4.0}{m}~(\mathrm{H})$, composed of two drift regions of \SI{2}{m}, with a central cathode, and two wire readout planes, as shown in the diagram in Fig.~\ref{fig:sbnd} (top left). The drift direction is perpendicular to the neutrino beam and  the maximal drift length (distance between the cathode and the wire planes) corresponds to about \SI{1.3}{ms} drift time for the nominal drift field of \SI{500}{V/cm}. Each drift volume wire readout is built from two interconnected Anode Plane Assemblies (APAs). Each APA consists of a $\SI{4.0}{m}\times\SI{2.5}{m}$ steel frame supporting three planes of \SI{150}{\mu m} copper-beryllium wires at pitch and plane spacing of \SI{3}{mm}.
Figure~\ref{fig:sbnd} (right) shows one completed APA, with a vertical wire collection plane, and two induction planes at $\pm$60$^{\circ}$ angles to the vertical. 
By appropriate voltage biasing, the first two Induction wire planes facing the drift region provide signals in a non-destructive way and the charge is collected on the last Collection wire plane. The electronics readout is composed of custom pre-amplifiers, commercial ADCs, and motherboards with onboard FPGA connected to the end of each wire plane and operating in the liquid argon. There is a jumpered interconnect between the two neighboring APA frames so they function as a single wire plane unit and signals can be collected at the outer edges of the TPC volume. The central cathode is a welded, electropolished assembly composed of a stainless steel tube frame supporting stainless mesh panels. The cathode will be biased at \SI{-100}{kV}. A coaxial high voltage feedthrough composed of a stainless steel core and grounding sheath with polyethylene (PE) insulator brings this bias into the LAr cryostat and contacts the cathode donut with a spring-loaded tip. The \sbnd field cage is composed of roll formed aluminium profiles, with PE end caps. This is identical to the field cage designed for DUNE~\cite{Abi:2018alz} and used in the DUNE prototype at CERN, ProtoDUNE~\cite{Abi:2017aow}.

\sbnd has a composite photon detector system that both enhances the amount of light collected and provides an R\&D opportunity for scintillation detection in LAr.  The primary system is an array of 120 8" Hamamatsu R5912-mod photomultiplier tubes (PMTs) mounted behind the TPC wire planes.  96 are coated in wavelength shifting tetraphenyl butadeine (TPB) and 24 are uncoated for directly observing visible light. In addition, two other technologies are used in the \sbnd light collection system, the ARAPUCA photon trap and light guiding acrylic bars. The ARAPUCA~\cite{Machado:2016jqe} (and X-ARAPUCA~\cite{Machado:2018rfb}) is a novel photon collection device composed of dichroic filter windows on a highly internally reflective box instrumented with silicon photomultipliers (SiPMs).  SBND will contain 8 of each type of ARAPUCA.  The light guide bars are composed of wavelength shifter (TPB) coated acrylic strips, each read out by an array of SiPMs on both ends.  In addition, highly reflective polymeric foils that have been  evaporatively coated with TPB and laminated to a thin substrate are sandwiched between the layers of conductive mesh in the cathode plane.  This specular reflective surface greatly enhances the uniformity of light collection in the \sbnd volume, and the total light yield of the system is as much as 100 photoelectrons per MeV of energy loss in the detector. 

The entire TPC is housed in a stainless steel membrane cryostat, which is a similar design and serves as a prototype for the cryostat of the DUNE experiment. The \sbnd TPC is supported from the cryostat roof, which contains the feedthroughs for all detector cables and the high voltage. Figure~\ref{fig:sbnd} (bottom left) shows the TPC attached to the cryostat lid before being inserted into the cryostat.

As \sbnd is located on the surface, in order to mitigate the cosmic ray background in the detector, it is surrounded on all sides by planes of extruded scintillator strips also read out by SiPMs, which act as a cosmic muon tracker.  In addition, the building is designed to support \SI{3}{m} of concrete overburden directly above the \sbnd detector.

\sbnd plays a role in the on-going R\&D effort to develop the LArTPC technology, testing several technologies that can be used in a future kiloton-scale neutrino detector for a long-baseline experiment. Major components for the experiment have been fabricated at different institutions in Europe, South America and the United States, and are arriving at Fermilab from around the world to be integrated into the detector. \sbnd is scheduled to begin commissioning in late 2020.

%%%%%%%%%%%%%%%%%%%%%%%%%%%%%%%%%%%%%%%%%%%%%%%%%%%%%%%%%%%
\subsection{The SBN Far Detector: ICARUS-T600}
\label{sec:icarus}
%%%%%%%%%%%%%%%%%%%%%%%%%%%%%%%%%%%%%%%%%%%%%%%%%%%%%%%%%%%

\begin{figure}
\centering
\begin{minipage}{.49\textwidth}
  \centering
  \includegraphics[width=1\textwidth]{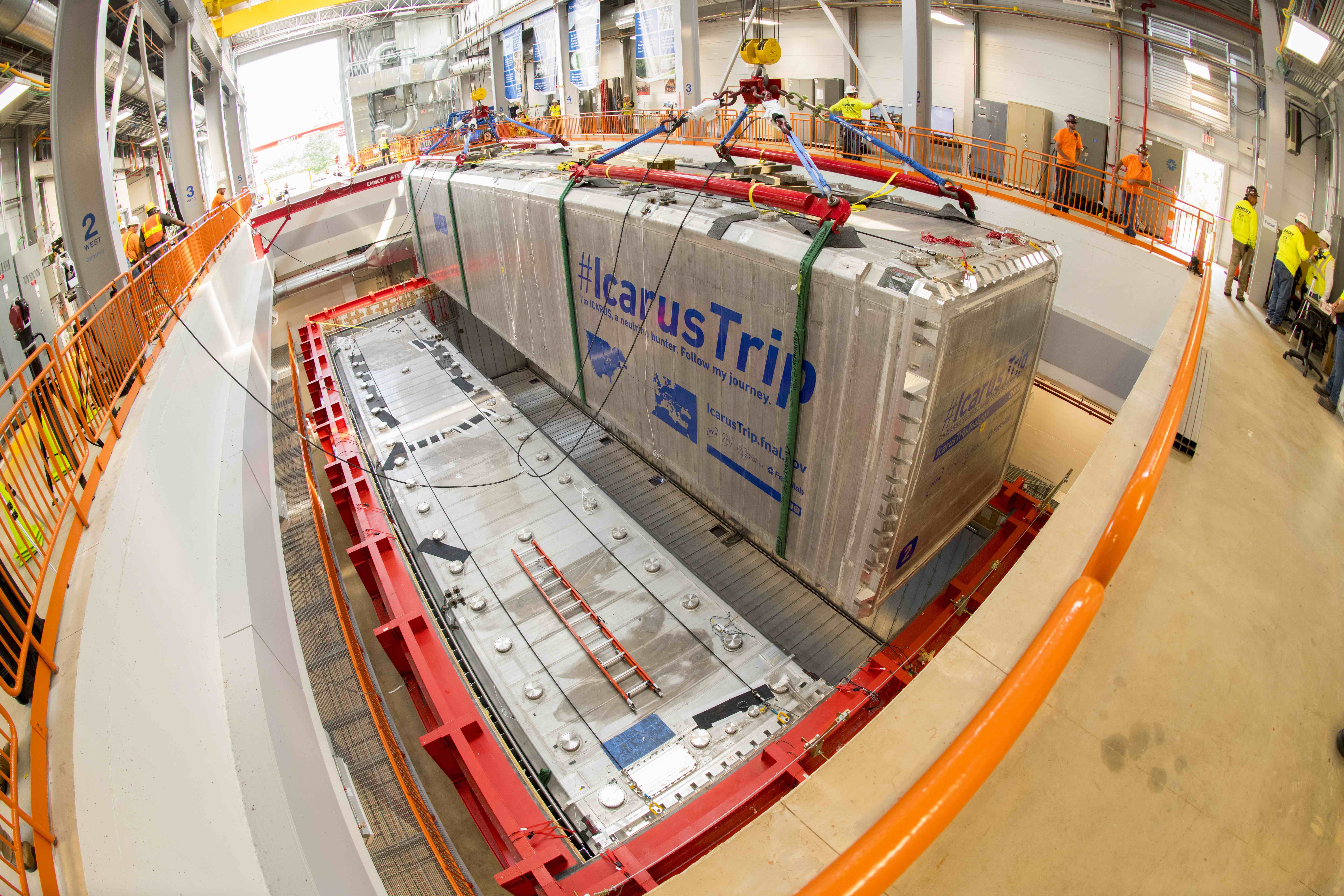}
\end{minipage}
\begin{minipage}{.49\textwidth}
  \centering
  \includegraphics[width=1\textwidth]{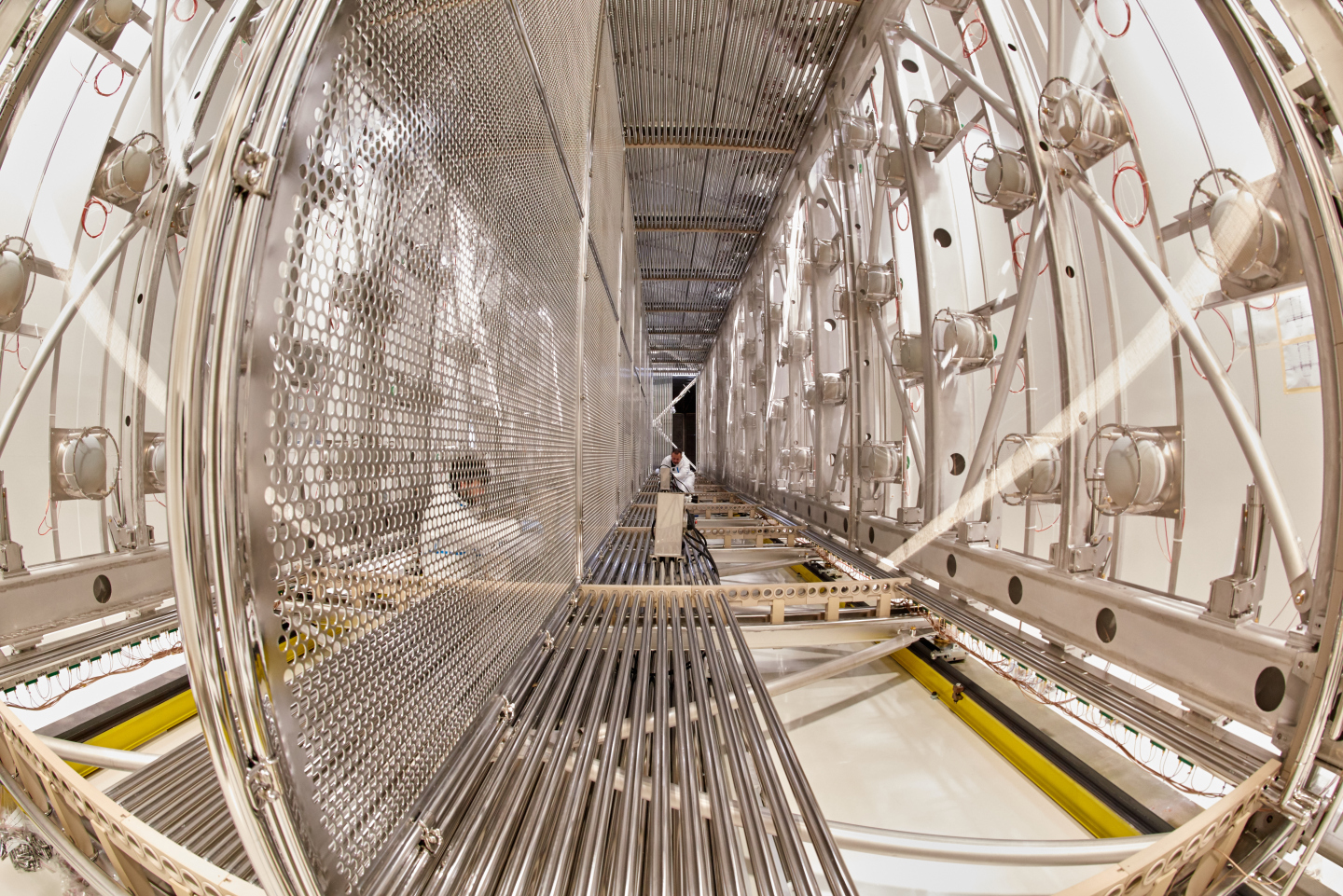}
\end{minipage}
\caption{{\it Left:}~The SBN far detector, the ICARUS-T600, being lowered into place in the far detector experimental hall at Fermilab in August 2018. Image credit:~Reidar Hahn, Fermilab. {\it Right:}~The inside of the detector during refurbishment at CERN showing the perforated cathode design and the \SI{8}{in}.~PMTs mounted behind the anode wires on both sides. Image credit: CERN.}
\label{fig:icarus}
\end{figure}

The ICARUS detector, a \SI{470}{ton} \lartpc, now part of the SBN program, was the culmination of a very successful R\&D campaign by the ICARUS collaboration to demonstrate the \lartpc technology for neutrino physics.  A detailed description of the ICARUS-T600 \lartpc detector can be found in Ref.~\cite{Amerio:2004ze}.  The  detector has been operated for 3 years (2011-2013) in the Gran Sasso Laboratory in Italy, where it demonstrated the underground operation of a large high-purity LAr detctor~\cite{Antonello:2015zea,Antonello:2014eha,Antonello:2014rwa}, measured the neutrino velocity~\cite{Antonello:2012be,Antonello:2012hg,ICARUS:2011aa}, and searched for evidence of neutrino oscillations~\cite{Antonello:2012pq} using the CERN Neutrino to Gran Sasso (CNGS) neutrino beam.   The detector is divided into two identical, adjacent cryostat modules (T300), as can be seen in Fig.~\ref{fig:icarus}.  Each T300 module houses two TPCs made of three parallel wire planes, \SI{3}{mm} apart, the first with horizontal wires and the other two at $\pm$60$^{\circ}$ from the horizontal direction (see Fig.~\ref{fig:icarus}, right). Wires are made of AISI 304V stainless steel with a wire diameter of 150 $\mu$m. The dimensions of the active volume in each T300 half module are $\SI{18.0}{m}~(\mathrm{L})\times\SI{3.2}{m}~(\mathrm{H})\times\SI{3.0}{m}~(\mathrm{W})$.  The two TPCs in each module are separated by a common cathode made of a stainless steel frame structure supporting punched stainless-steel sheets. The maximal drift length (distance between the cathode and the wire planes), is \SI{1.5}{m} corresponding to about \SI{1}{ms} drift time for the nominal drift field of \SI{500}{V/cm}. The readout electronics are located outside the detector. A Decoupling Board receives the signals from the chamber and passes them to an Analogue Board via decoupling capacitors; it also provides wire biasing voltage and the distribution of the test signals.
The Analogue Board hosts the front-end amplifier and performs digitization. 

In 2014 the ICARUS detector was transported to CERN and underwent a significant overhauling. The most important upgrades were: higher-performance TPC readout electronics, the installation of a new scintillation light detection system, the construction of new cold vessels, and the refurbishment of the cryogenics and LAr purification systems. The new TPC readout electronics is hosted in a very compact set-up, with crates mounted directly on top of the signal flanges. The new scintillation light detection system is composed of 360 8" Hamamatsu R5912-mod PMTs, same as those used in SBND. 
The detector overhauling process has been concluded with the installation of the two internal detectors, completely refurbished, in the new aluminum vessels at CERN. The two ICARUS modules have then been transported to Fermilab in July 2017. The final placement of the detectors in the SBN far detector building at Fermilab was completed in August 2018 (see Fig.~\ref{fig:icarus}, left) and cryogenic plant and detector readout installation are now in progress.  Similarly to the near detector, a segmented cosmic ray tagging system composed of plastic scintillation slabs read out by SiPMs will surround the detector in order to mitigate cosmic generated backgrounds. Commissioning of the ICARUS detector at Fermilab is planned for fall 2019.

%%%%%%%%%%%%%%%%%%%%%%%%%%%%%%%%%%%%%%%%%%%%%%%%%%%%%%%%%%%%%%%%%%%%%%%%%%%%%%
\section{SCIENCE OF THE SBN PROGRAM}
\label{sec:science}
%%%%%%%%%%%%%%%%%%%%%%%%%%%%%%%%%%%%%%%%%%%%%%%%%%%%%%%%%%%%%%%%%%%%%%%%%%%%%%

The Short-Baseline Neutrino program has been designed specifically to address the sterile neutrino interpretation of the experimental anomalies discussed in Section~\ref{sec:steriles}. However, the science capabilities of SBN extend well beyond the flagship oscillation searches.  Millions of neutrino interactions will be recorded on argon in high precision detectors, enabling a broad research program in neutrino-argon scattering that has direct relevance for plans to use liquid argon detectors in future neutrino experiments, especially DUNE.  In addition, as the world-wide effort to discover new physics beyond the Standard Model continues, the SBN program has received attention from experimenters and theorists alike for the opportunity it presents to contribute to this quest. In this section, we summarize these three pillars of the SBN physics program.    

\subsection{Oscillation Searches and Sterile Neutrino Sensitivity} 
\label{sec:osc}

\begin{figure}
    \begin{minipage}{0.5\textwidth}
        \centering
        \includegraphics[width=0.97\textwidth,trim={4mm 1mm 3mm 4mm},clip]{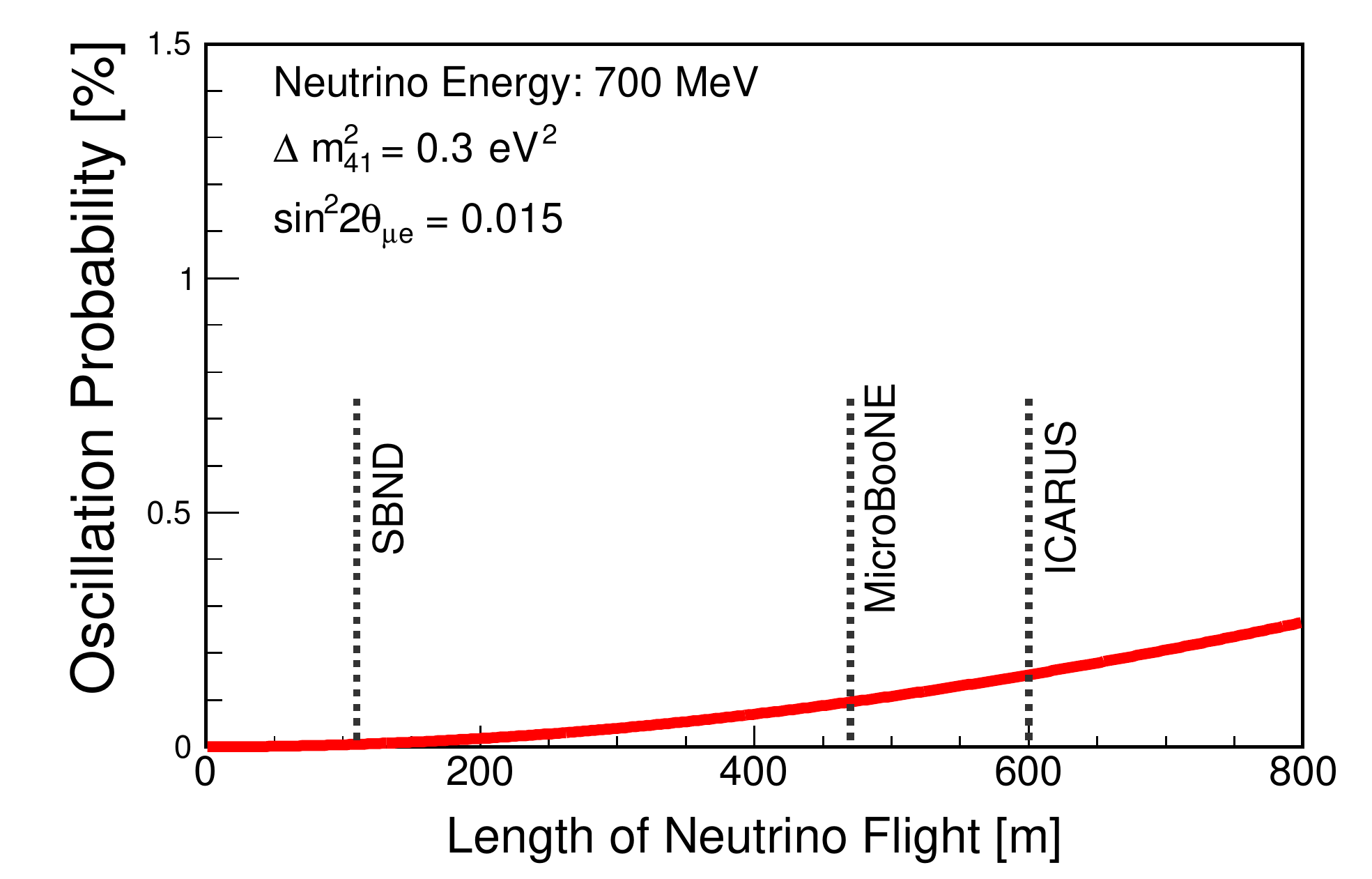}
    \end{minipage}
    \begin{minipage}{0.5\textwidth}
        \centering
        \includegraphics[width=0.97\textwidth,trim={4mm 1mm 3mm 4mm},clip]{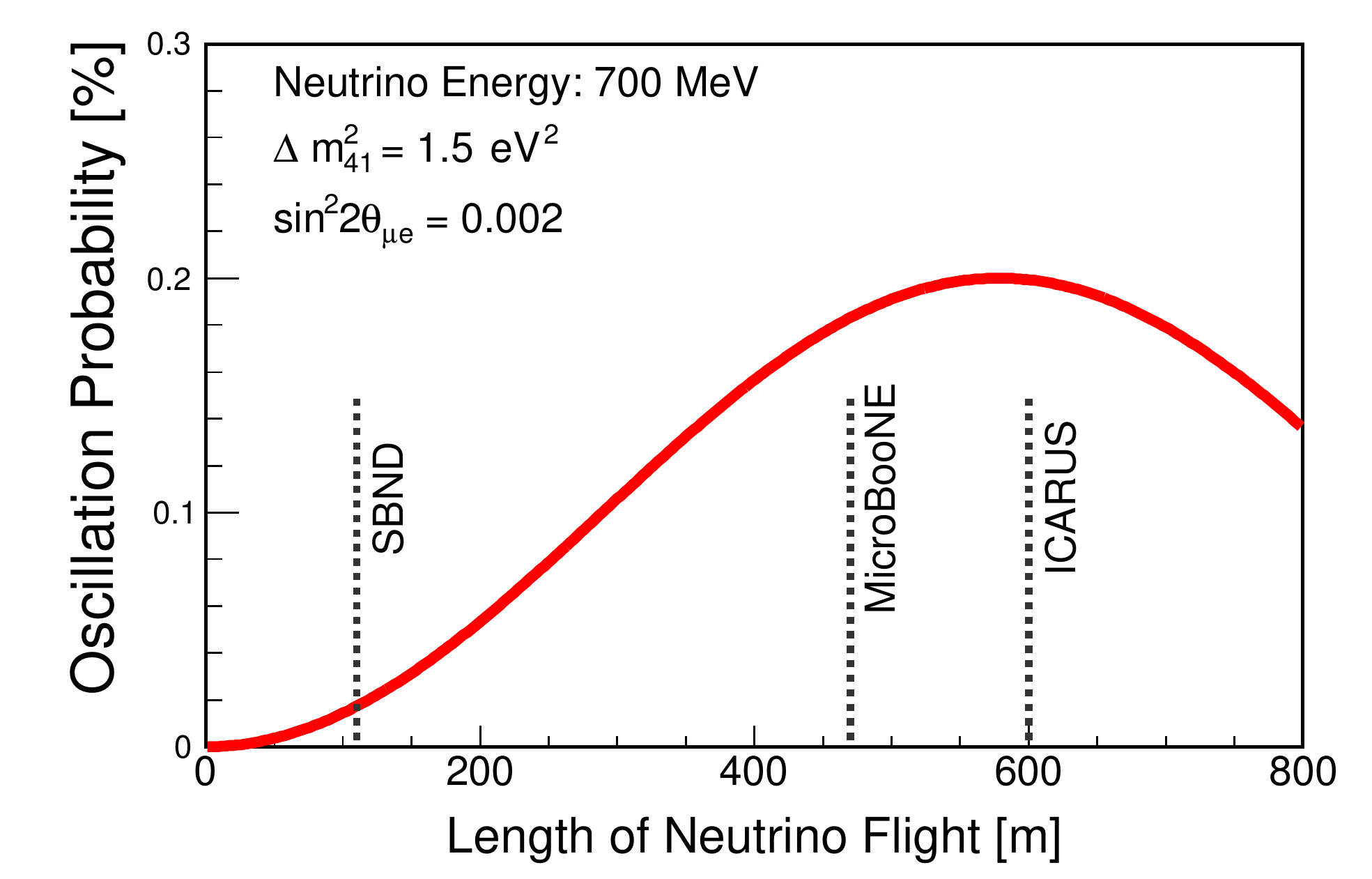}
    \end{minipage}
    \begin{minipage}{0.5\textwidth}
        \centering
        \includegraphics[width=0.97\textwidth,trim={4mm 1mm 3mm 4mm},clip]{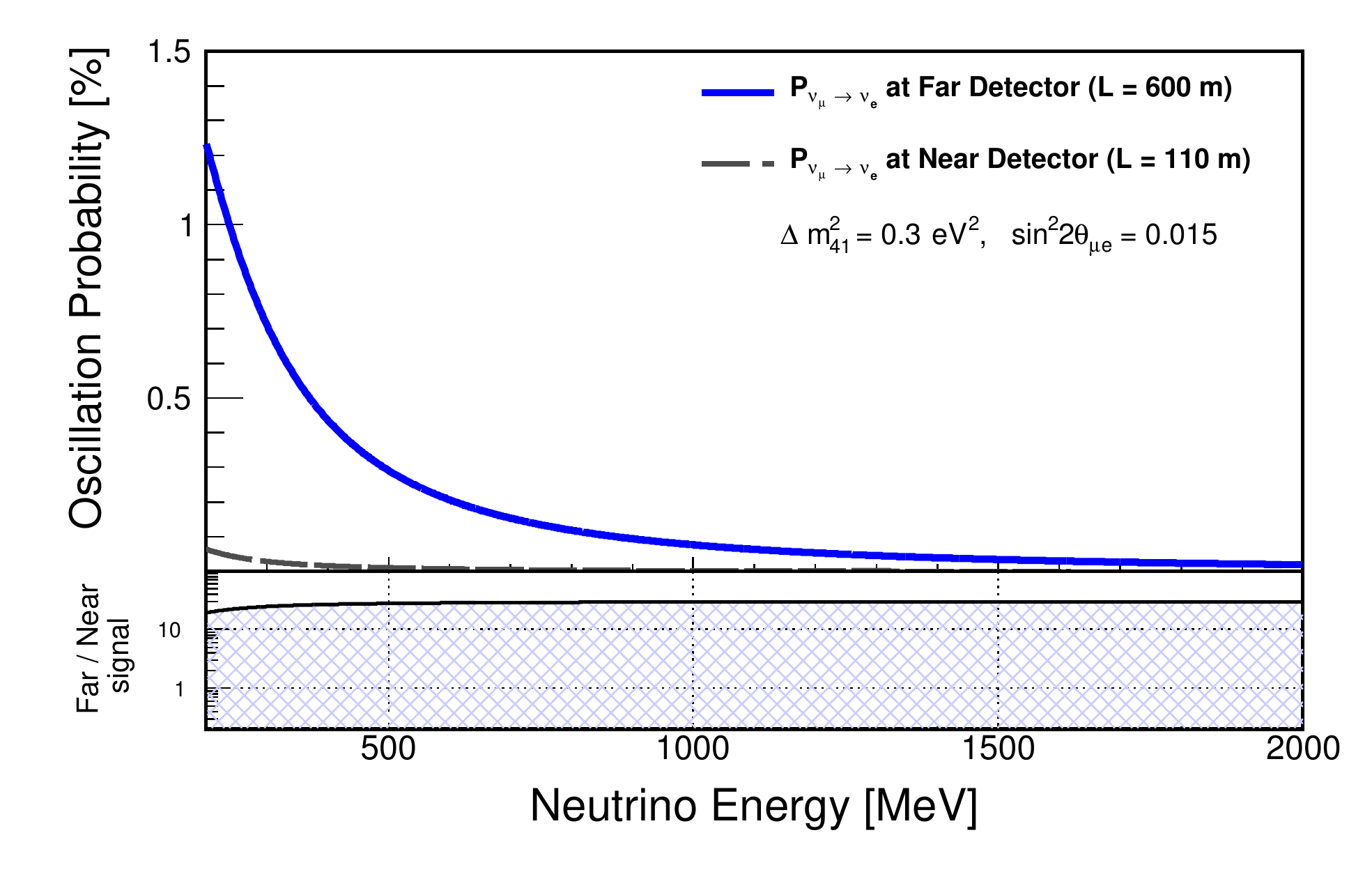}
    \end{minipage}
    \begin{minipage}{0.5\textwidth}
        \centering
        \includegraphics[width=0.97\textwidth,trim={4mm 1mm 3mm 4mm},clip]{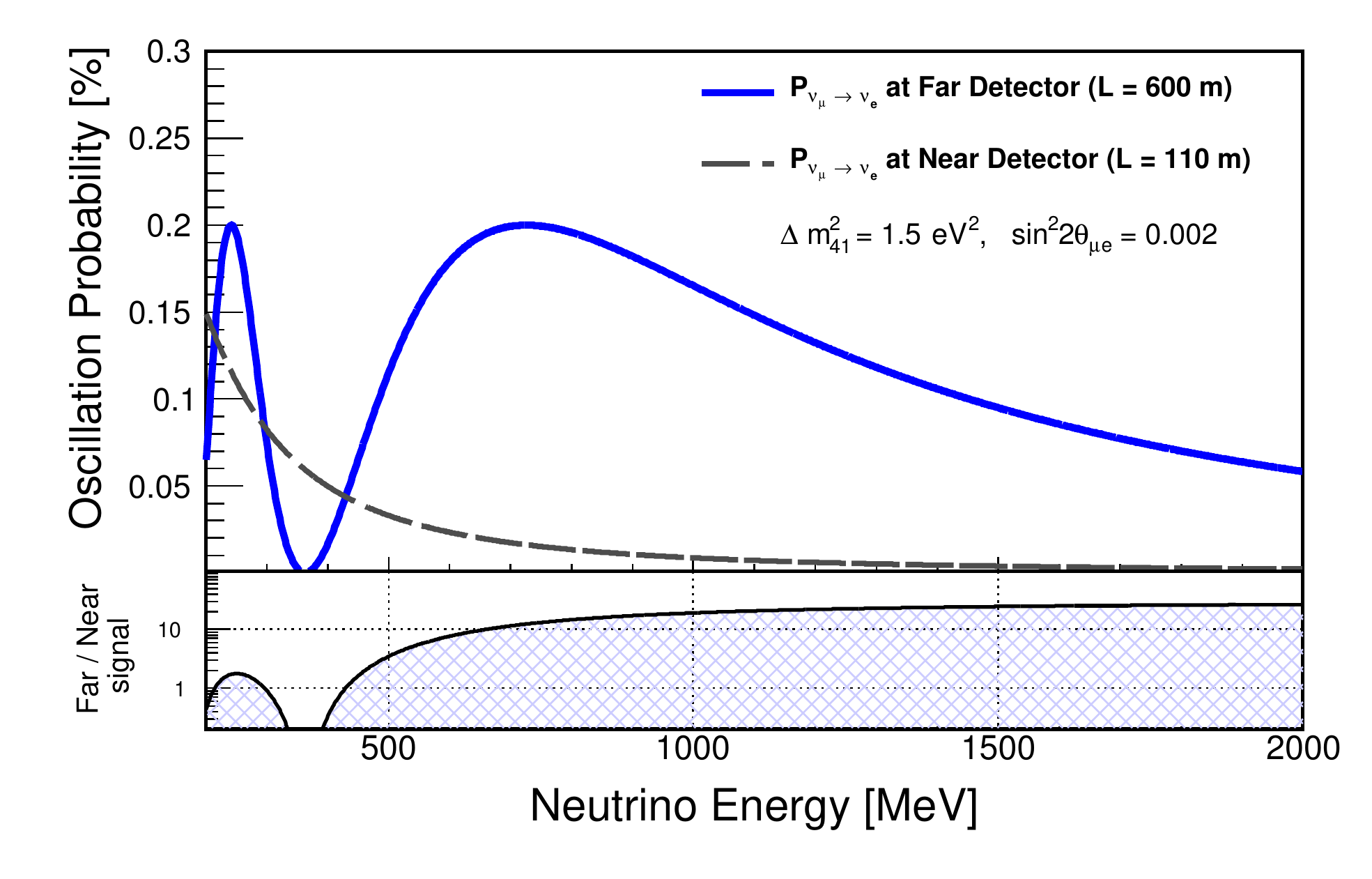}
    \end{minipage}
    \caption{{\it Upper panels:}~$\nu_\mu\to\nu_e$ oscillation probability for a \SI{700}{MeV} neutrino as a function of the baseline for two different benchmark points in a 3+1 sterile neutrino scenario. {\it Lower panels:}~$\nu_\mu\to\nu_e$ oscillation probabilities, at \SI{110}{m} and \SI{600}{m}, as a function of the neutrino energy for the same benchmark points. The far-over-near ratio of appearance probabilities is also shown.}
    \label{fig:sbn-oscillations}
\end{figure}

SBN has several advantages over previous experiments in the search for $\sim$\SI{1}{eV} sterile neutrinos. In particular, the multi-detector design is essential to achieving SBN's world-leading sensitivity to short-baseline neutrino oscillations. The locations of the near and far detectors are optimized for maximal sensitivity in the most relevant ranges of oscillation parameters. As summarized in Section~\ref{sec:landscape} and seen in Fig.~\ref{fig:tension}, the global $\nu_e$ appearance data point to a mass splitting, $\Delta m^2_{41}$, between \SI{0.3}{eV^2} and \SI{1.5}{eV^2} with a mixing strength in the range $0.002\lesssim\sin^22\theta_{\mu e}\lesssim0.015$.  Figure~\ref{fig:sbn-oscillations} depicts the shape of the oscillation probability within SBN for sets of parameters spanning this range. In the top figures, the evolution of the electron neutrino appearance probability near the peak neutrino beam energy (\SI{700}{MeV}) is shown as a function of the travel distance of the neutrino. Below this are the oscillation probabilities versus neutrino energy at both the near ($L = \SI{110}{m}$) and far ($L = \SI{600}{m}$) detector locations. Oscillations are visible in the far detectors for all oscillation parameters in the range indicated by global analyses. For larger $\Delta m^2$, a small oscillation signal does begin to appear at the lowest neutrino energies in the near detector, but the very different shape and higher level of oscillation at most energies at the far detector preserves the strong sensitivity of the experiment up to several eV$^2$.  

The near-far detector combination is crucial because, as in most modern oscillation experiments, it allows for optimal control of systematic uncertainties in the search for oscillation signals.  Precision oscillation studies in a single detector experiment~\cite{Aguilar-Arevalo:2018gpe} or even in multi-detector experiments with different near and far detector technologies~\cite{Cheng:2012yy,Abe:2018wpn} are severely challenged by uncertainties in the neutrino flux and/or the modeling of neutrino-nucleus scattering.  Uncertainties in the absolute neutrino fluxes and interaction cross sections at BNB energies are also large (10-30\%), but the highly correlated event rates in the SBN near and far detectors (since they utilize the same interaction target medium and detection technique) will enable a significant cancellation of the flux and cross section uncertainties when comparing between the two.     

A second key advantage of SBN is the ability of the liquid argon TPC technology to reduce the main backgrounds that affected the MiniBooNE experiment, one of the main anomalies to be addressed. Specifically, photon-induced electromagnetic showers in the MiniBooNE Cherenkov detector are indistinguishable from those induced by an electron, as from charged-current $\nu_e$ scattering.  The fine sampling calorimetry of the \lartpc provides multiple handles to separate electron and gamma induced activity.  First, a photon will propagate some distance before interacting ($X_0 = \SI{14}{cm}$ in LAr), so if a vertex can be identified, then the gap between the vertex and the start of an electromagnetic shower is a clear photon signature.  Second, when the photon converts to produce an $e^+e^-$ pair, the resulting ionization in the first few centimeters will be consistent with two minimum ionizing particles (mips), distinguishing it from the single mip deposit of an electron \cite{Acciarri:2016sli}.

In 2014, the scientific collaborations representing the three detectors of SBN (SBND, \uboone, and ICARUS) came together to do a comprehensive analysis of the physics capabilities of SBN, which were summarized in the program proposal~\cite{Antonello:2015lea}.  The analysis leveraged several advanced simulation tools, including the robust simulation of the Booster Neutrino Beam developed by the MiniBooNE collaboration~\cite{AguilarArevalo:2008yp}, the GENIE neutrino interaction event generator and built-in systematic error machinery~\cite{Andreopoulos:2009rq,Andreopoulos:2015wxa}, and the GEANT4~\cite{Agostinelli:2002hh} and FLUKA~\cite{BOHLEN2014211,Ferrari:2005zk} particle transport and interaction codes.  Correlations in the impact of neutrino flux and neutrino-argon interaction model variations were quantified using these tools and applied in the sensitivity analysis.  

With neutrino flux and interaction uncertainties minimized through the strong detector-to-detector correlations, the emphasis turns to controlling systematic differences in the selection and reconstruction of neutrino events.  Detector uncertainties were estimated at the time of the proposal, though with less sophisticated tools, and this remains a major focus of current analysis efforts. In particular, much is now known from the experience operating the \uboone detector since late in 2015. Significant advances have been made in TPC data analysis using \uboone data, including TPC noise filtering~\cite{Acciarri:2017sde}, wire signal processing and deconvolution~\cite{Adams:2018dra,Adams:2018gbi}, reconstruction algorithm development~\cite{Acciarri:2016ryt,Acciarri:2017hat,Adams:2018bvi}, and the simulation of bulk properties such as the `space charge effect', which is the distortion of the electric drift field caused by the buildup of positive ions in the LAr from the high rate of cosmic ray activity in detectors that are operated near the surface.     

Finally, a critical outcome of the proposal sensitivity analysis was the recognition of the importance of including multiple handles to mitigate cosmic induced backgrounds in the analysis.  The \lartpc is an intrinsically slow technology, with drift times in the millisecond range, so detectors at the surface record significant cosmic activity with each readout (5-15 cosmic muons are seen per readout in the case of the SBN detectors).  The SBN near and far detectors, therefore, have been designed to include external cosmic ray tagger detector systems (CRTs) with nearly $4\pi$ coverage and based on solid scintillator technology to achieve hit timing resolution of a few nanoseconds.  In addition, both the near and far detector buildings are designed to support \SI{3}{meters} of concrete overburden directly above the pits where the detectors are installed.   This shielding will absorb more than 99\% of the photon and hadron content of cosmic showers hitting the experimental halls. The shielding and CRTs provide a powerful combination for cosmic background mitigation that is essential to the physics goals of SBN.

\begin{figure}
  \checkoddpage
  \edef\side{\ifoddpage l\else r\fi}%
  \makebox[\textwidth][\side]{%
\begin{minipage}{.58\textwidth}
  \centering
  \includegraphics[width=1\linewidth,trim={6mm 2mm 26mm 13mm},clip]{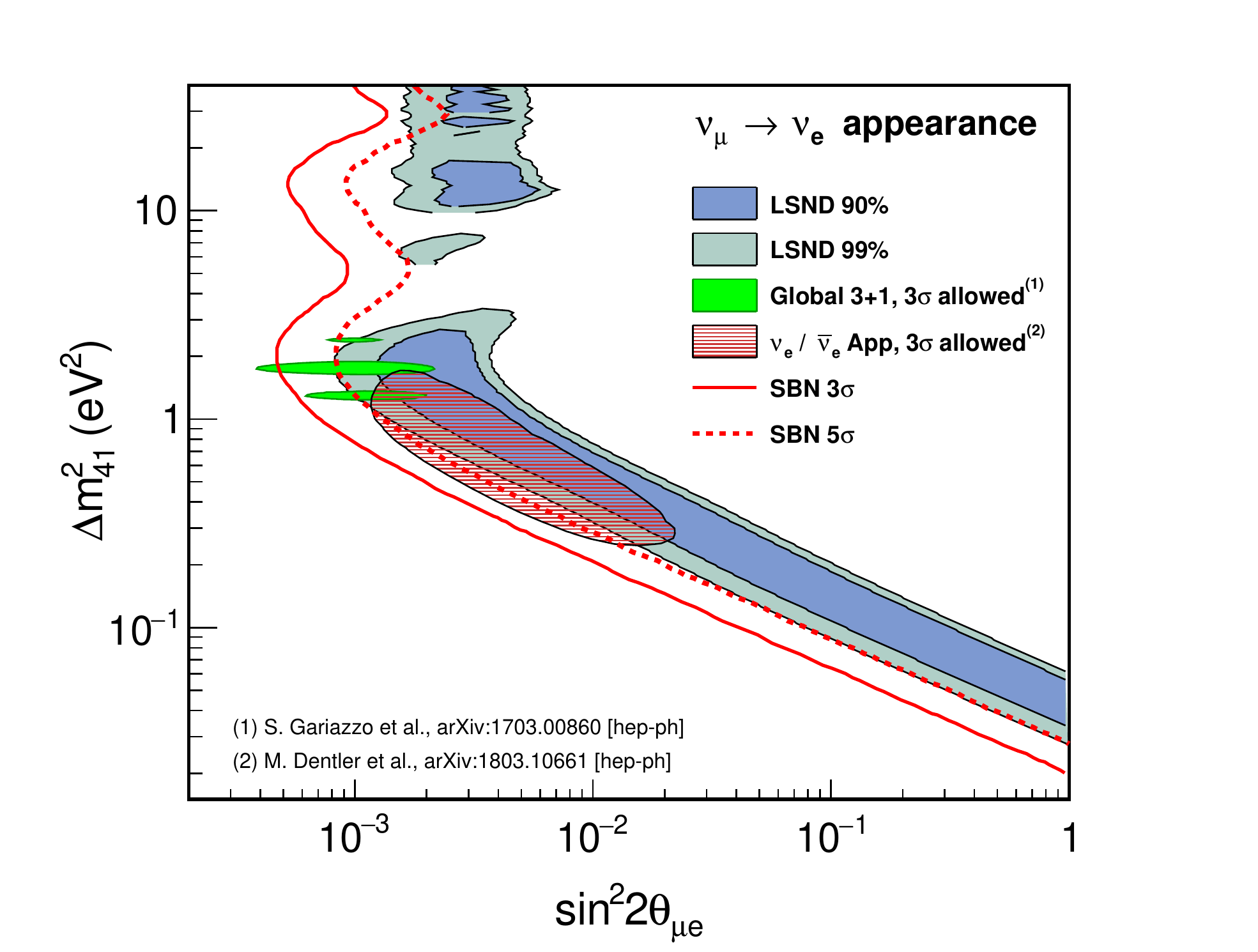}
\end{minipage}%
\begin{minipage}{.58\textwidth}
  \centering
  \includegraphics[width=1\linewidth,trim={6mm 2mm 26mm 13mm},clip]{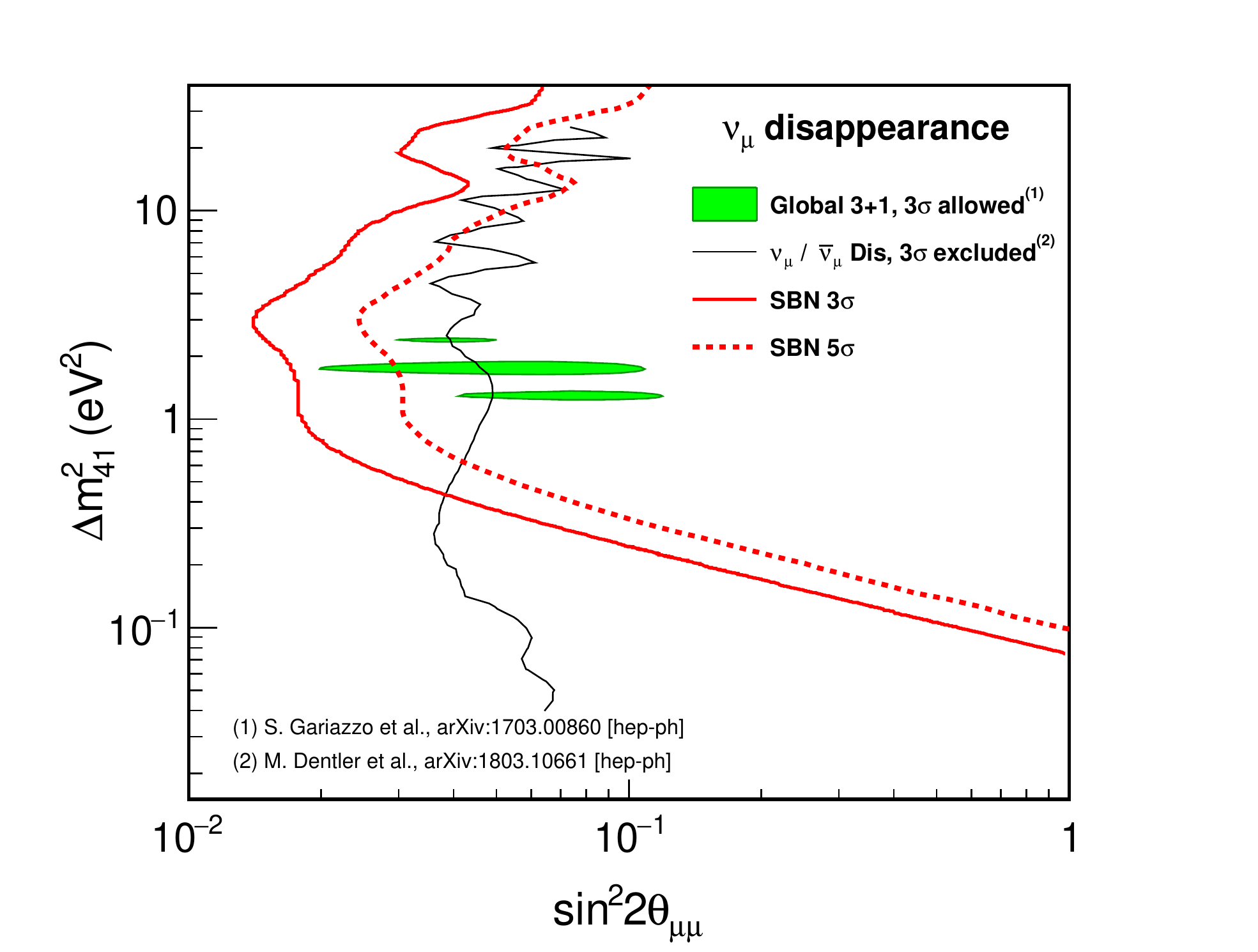}
\end{minipage}%
}%

  \checkoddpage
  \edef\side{\ifoddpage l\else r\fi}%
  \makebox[\textwidth][\side]{%
\begin{minipage}{1.15\textwidth}
\caption{SBN  $3\sigma$ (solid red line) and $5\sigma$ (dotted red line) sensitivities to a light sterile neutrino in the $\nu_\mu\to\nu_e$ appearance channel {\it (left)} and $\nu_\mu\to\nu_\mu$ disappearance channel {\it (right)}. For comparison, the LSND preferred region at 90\% C.L. (shaded blue) and 99\% C.L. (shaded gray) is presented~\cite{Aguilar:2001ty}. Moreover, the global $\nu_e$ appearance (shaded red) and global $\nu_\mu$ disappearance (black line) $3\sigma$ regions from Ref.~\cite{Dentler:2018sju} are also included. Finally, the $3\sigma$ global best fit regions from Ref.~\cite{Gariazzo:2017fdh} are shown in green. The sensitivities are reproduced from the SBN proposal~\cite{Antonello:2015lea}.}
\label{fig:sensitivity}
\end{minipage}%
}%
\end{figure}

The projected sensitivities to $\nu_\mu \to \nu_e$ conversion and $\nu_\mu$ disappearance oscillation signals are shown in Fig.~\ref{fig:sensitivity}.  The analysis is presented in the context of a 3+1 sterile neutrino model according to Eqs.~\ref{eq:APP} and \ref{eq:DIS-mu}.  Event rates and systematic uncertainties and their correlations were determined from the full BNB and GENIE simulation codes, as described above. An uncorrelated detector systematic uncertainty at the level of 3\% is assumed. Statistical errors are derived assuming an exposure of $6.6\times10^{20}$ protons delivered to the BNB target, which corresponds to approximately three years of operation, for both the near and far detectors.  

The capability to search for evidence of oscillation through muon neutrino disappearance is a very important feature of the SBN program, owing to the intense muon neutrino beam and multiple detectors.  The severe tension between existing $\nu_e$ appearance and $\nu_\mu$ disappearance data, as discussed in Section~\ref{sec:steriles}, presents a major challenge to the sterile neutrino interpretation at present.  The observation of muon neutrino disappearance, commensurate with an appearance signal, would be essential to the interpretation of any electron neutrino excess as being due to the existence of sterile neutrinos.  

Focusing on a 3+1 sterile neutrino scenario, the $3\sigma$ and $5\sigma$ sensitivities of SBN are presented as solid and dotted red lines, respectively, in Fig.~\ref{fig:sensitivity}. To put the SBN sensitivity into perspective, several related results are superimposed for comparison.  To start with, SBN was designed to cover, at $\geq5\sigma$, the full 99\% allowed region of the original LSND appearance result reported in 2001~\cite{Aguilar:2001ty} in the ($\sin^2 2\theta_{\mu e},\Delta m^2_{41}$) plane.  This is presented as the blue (90\% C.L.) and gray (99\% C.L.) regions in the left panel. 

Since then, only MiniBooNE has reported new anomalies in $\nu_e$ appearance, and limits from other experiments have reduced the possible sterile neutrino parameter space significantly over the years.  Besides, as discussed before, there is a strong tension between the appearance and disappearance data sets. Thus, to show how SBN sensitivity compares to each data set independently, we also  present in Fig.~\ref{fig:sensitivity} the preferred regions for all $\nu_\mu\to\nu_e$ appearance data alone (left panel, shaded red) and the limit imposed by all $\nu_\mu$ disappearance data alone (right panel, black solid line) at $3\sigma$ C.L. from Ref.~\cite{Dentler:2018sju}. Notice that SBN alone may be able to rule out almost all the global appearance preferred region at $5\sigma$, while the expected sensitivity on the $\nu_\mu$ disappearance channel is better than the global constraints for a large range of $\Delta m^2_{41}$.

Regardless of the tension between data sets, it is instructive to compare SBN sensitivities to parameter regions that are allowed in a global analysis of all signal and null results from all three oscillation channels, such as in Ref.~\cite{Gariazzo:2017fdh}. The $3\sigma$ allowed parameter regions are displayed as the shaded green regions in both panels of Fig.~\ref{fig:sensitivity}.  One sees that $\Delta m^2$ is tightly constrained around 1-\SI{2.5}{eV^2} in order to satisfy all data, and the severe tension between data sets is manifested as the stretching of the allowed regions in mixing strength.
Note that the preferred values of $\sin^22\theta_{\mu\mu}$ are necessarily accompanied by non-zero values of $\sin^22\theta_{ee}$ due to non-zero $\sin^22\theta_{\mu e}$, see Eqs.~(\ref{eq:APP}-\ref{eq:DIS-mu}) and Ref.~\cite{Gariazzo:2017fdh}, suggesting again the complementarity of the reactor neutrino experiments and SBN. 
Figure~\ref{fig:sensitivity} shows that SBN has its best sensitivity in the regions of remaining allowed parameter space, in both channels. It is exciting to see how SBN is primed to rule on the possibility of the existence of sterile neutrinos.

%%%%%%%%%%%%%%%%%%%%%%%%%%%%%%%%%%%%%%%%%%%%%%%%%%%%%%%%%%%%%%%%%%%%%%%%%%
\subsection{Neutrino-Nucleus Scattering} 
\label{sec:xsec}
%%%%%%%%%%%%%%%%%%%%%%%%%%%%%%%%%%%%%%%%%%%%%%%%%%%%%%%%%%%%%%%%%%%%%%%%%%

Neutrino-nucleus interactions are critical to understand in neutrino oscillation experiments~\cite{Alvarez-Ruso:2017oui,Mosel2018}, including the DUNE liquid argon long-baseline program. The SBN physics program includes the study of neutrino-argon cross sections with millions of interactions using the well characterized neutrino fluxes of the BNB. Being the near detector, \sbnd observes the largest flux of neutrinos of the three detectors and provides an ideal venue to conduct precision studies of the physics of neutrino-argon interactions in the GeV energy range. The experiment will collect enormous neutrino event samples and will make the world's highest statistics cross section measurements for many $\nu$-Ar scattering processes. In \sbnd, more than 2 million neutrino interactions will be collected per year in the full active volume (assuming 2.2$\times$10$^{20}$ protons on target), with 1.5 million $\nu_{\mu}$ charged-current (CC) events.  This will quickly reduce the statistical uncertainty to well below the percent level, making systematic errors dominant. In addition to the large number of $\nu_{\mu}$ events, there will be a high statistics $\nu_e$ sample, which will allow for both inclusive and exclusive measurements of electron neutrino interactions. SBND will record around 12,000 $\nu_e$ events per year. Measurements of this interaction will be extremely beneficial for both the SBN and DUNE physics programs. To put the \sbnd measurements into context of other LAr detectors, each year exposure of \sbnd will provide an event sample 6-7 times larger than the one available in the full \uboone phase I run.

\begin{figure}
\begin{minipage}{.5\textwidth}
  \centering
  \includegraphics[width=1\linewidth,trim={4mm 0mm 5mm 6mm},clip]{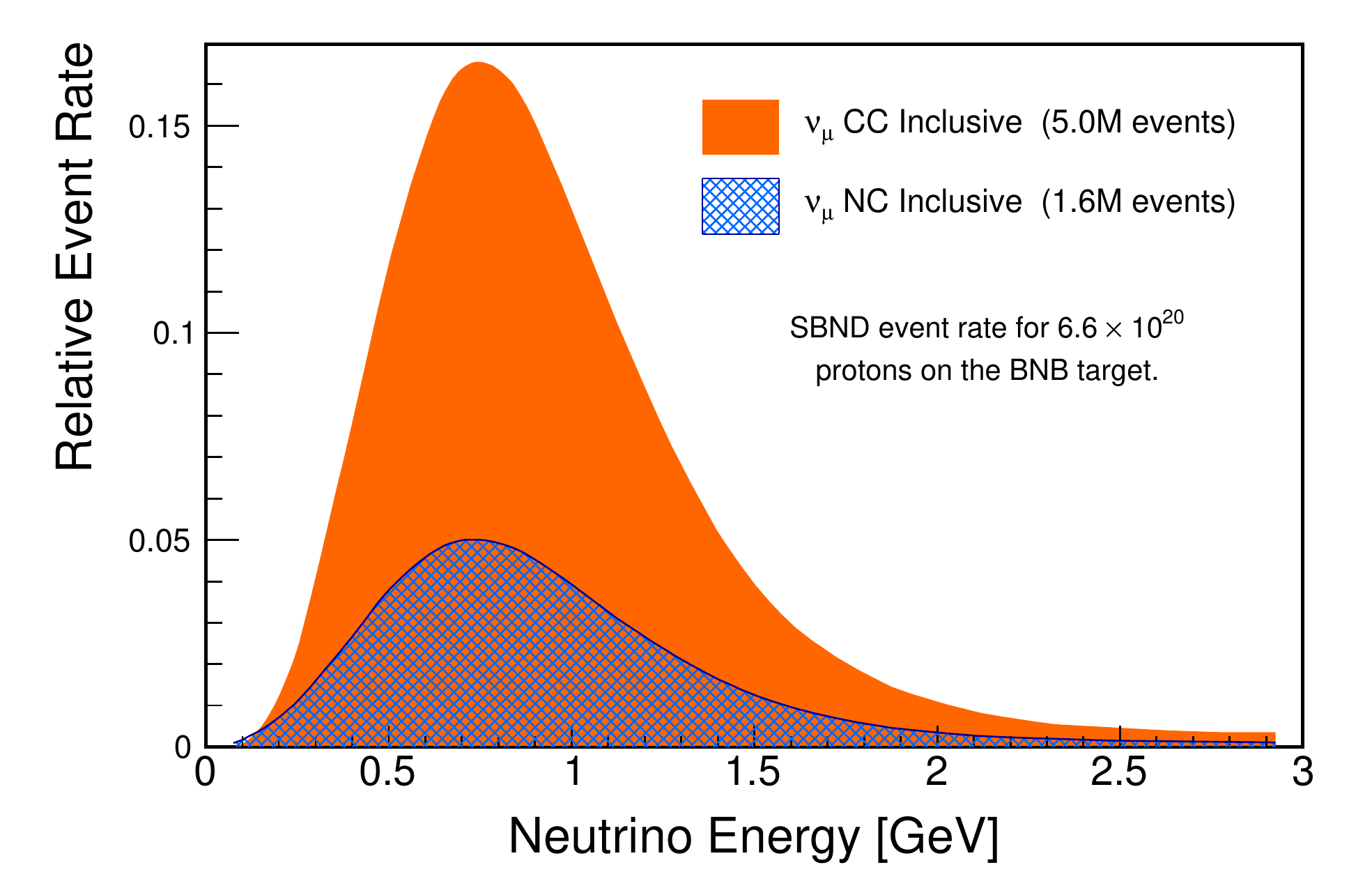}
\end{minipage}%
\begin{minipage}{.5\textwidth}
  \centering
  \includegraphics[width=1\linewidth,trim={4mm 0mm 5mm 6mm},clip]{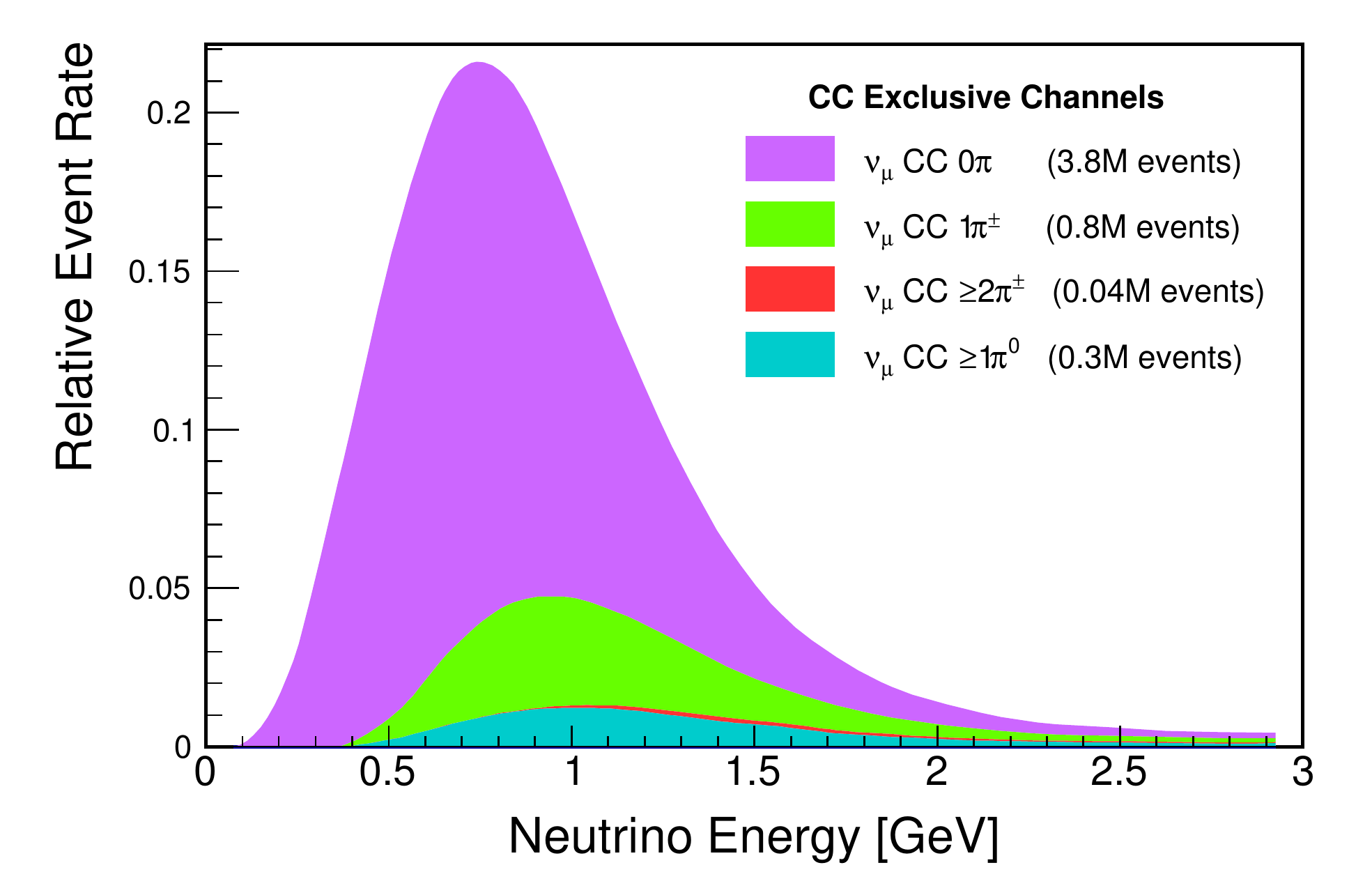}
\end{minipage}
\caption{Event spectra and rates in \sbnd for $6.6\times10^{20}$ protons on target ($\sim$3 years of operation). {\it Left:}~The total $\nu_\mu$ inclusive charged-current and neutral-current event spectra (shown not stacked). {\it Right:}~The exclusive channel breakdown of the $\nu_\mu$ charged-current sample discriminated according to the number of pions in the final state (shown stacked). The spectra are normalized to show relative rates, with the total events expected for the exposure indicated in the legend.}
\label{fig:xsec}
\end{figure}

Figure~\ref{fig:xsec} (left) shows the spectra of $\nu_{\mu}$ CC and neutral-current (NC) interactions and the total numbers of events expected for an exposure of $6.6\times10^{20}$ protons on target (POT). Note that the BNB neutrino flux spectrum at \sbnd peaks near the neutrino energy of the second oscillation maximum for DUNE (\SI{0.8}{GeV}) and includes a substantial sample up to the first DUNE oscillation maximum (\SI{2.6}{GeV}). 

\sbnd will perform many exclusive measurements of different final states for $\nu_{\mu}$ and $\nu_e$ events with high precision and will measure nuclear effects from the comparison with different Monte Carlo (MC) simulations.  Figure~\ref{fig:bsm} (right) shows the expected rates of $\nu_{\mu}$ CC events separated into their main experimental topologies for the same $6.6\times10^{20}$ POT exposure. The largest event sample corresponds to the $\nu_{\mu}$ charged-current ``0 meson'' final state, where there is an outgoing muon, one or more recoil nucleons, and no outgoing pions or kaons. This cross section for scattering off nuclei largely depends on final state interactions and other nuclear effects and \sbnd data will allow the study of nuclear effects in neutrino interactions in argon nuclei with high precision. 
This data will inform neutrino Monte Carlo generators and aid in disentangling neutrino-nuclear interaction phenomenology by discriminating between final state interaction models. One example of the statistical power of the data is in measurements of neutrinos scattering off correlated nucleon pairs -- according to current simulations there will be $\sim$360,000 events per year with one muon and two protons ($1\mu + 2p$) in the final state. 

The high interaction rate will also allow \sbnd to measure interaction channels which remain unmeasured on argon. There are many rare interaction channels which can be probed by \sbnd, for example production of hyperons $\Lambda^0$ and $\Sigma^+$, for which \sbnd will collect a total data set of several thousand events over 3 years, recording more than the current historical data set each month. \sbnd will also see $\sim$400 $\nu e \rightarrow \nu e$ elastic scattering events in $6.6\times10^{20}$ POT. These events provide a unique topological signature, a very forward electron with no activity around the vertex, easily identified in a \lartpc. The elastic scattering of neutrinos on electrons is a process with a well known theoretical cross section, and with this event sample a measurement of the neutrino flux can be made~\cite{Park:2015eqa}. 

Furthermore, the MicroBooNE and ICARUS detectors are  respectively located approximately 8$^\circ$ and 6$^\circ$ off-axis to the higher energy NuMI beam, produced by \SI{120}{GeV} protons from the Fermilab Main Injector directed onto a carbon target~\cite{Adamson:2015dkw,Aliaga:2016oaz}. MicroBooNE and ICARUS can also study neutrino-argon cross sections exploiting the NuMI beam. 
ICARUS will collect a large neutrino event sample in the 0-\SI{3}{GeV} energy range with an enriched component of electron neutrinos ($\sim$5\%).  Muon neutrino event rates in the T600 from NuMI are comparable with the ones from the BNB, while the electron neutrino component is enhanced by an order of magnitude in the off-axis beam from the dominant three body decay of secondary kaons. ICARUS will see about 100,000 $\nu_{\mu}$ and 10,000 $\nu_e$ NuMI off-axis events per year.

All together, SBN data have the potential to transform our understanding of the physics of low-energy neutrino-nucleus scattering and will be the key input to refining the modeling of $\nu$-Ar interactions, in particular, before the DUNE era.

%%%%%%%%%%%%%%%%%%%%%%%%%%%%%%%%%%%%%%%%%%%%%%%%%%%%%%%%%%%%%%%%%%%%%%
\subsection{New Physics Opportunities at SBN}
\label{sec:bsm}
%%%%%%%%%%%%%%%%%%%%%%%%%%%%%%%%%%%%%%%%%%%%%%%%%%%%%%%%%%%%%%%%%%%%%%

Liquid argon TPC technology, with unprecedented event reconstruction, excellent particle identification, and fine-sampling calorimetry opens up invaluable opportunities for new physics searches.  A high intensity beam leading to large statistics will allow for excellent sensitivity to weakly coupled physics.   Below we present a summary of new physics scenarios that can potentially be probed at SBN besides the light sterile neutrino framework aforementioned, with a brief description of them and the important signatures in SBN. 
For each new physics scenario the searches will need to be carefully optimized for the SBN detectors, taking into account the challenge of substantial cosmic activity in \lartpcs operating near the surface.  In scenarios where new states are produced in the target (millicharged particles, light dark matter, etc), the Booster Neutrino Beam energy profile will play a crucial role in the experimental sensitivity.  We will see that new physics observables can generically be placed into two categories: modifications to oscillation physics and novel experimental signatures. 

\vspace{0.5em}

{\bf Electronvolt scale sterile neutrinos decaying} to active neutrinos and a Majoron or gauge boson would lead to new features in the active neutrino energy spectrum with respect to 3+N scenarios, significantly changing the sensitivity of some experiments to sterile neutrinos and providing a possible explanation of the LSND/MiniBooNE anomalies~\cite{Bai:2015ztj, Moss:2017pur}. 

{\bf Large extra dimensions.} The smallness of neutrino masses could be explained by the presence of large flat extra dimensions in which right-handed neutrinos propagate in the $n$-dimensional bulk, while the Standard Model is confined to the brane~\cite{ArkaniHamed:1998rs, Antoniadis:1998ig, ArkaniHamed:1998nn}. This model gives rise to a Kaluza-Klein tower of sterile neutrinos which mix with active neutrinos and thus can induce short-baseline oscillations~\cite{Mohapatra:2000wn, Davoudiasl:2002fq, Machado:2011jt} and possibly explain the reactor anomaly~\cite{Machado:2011kt}. Both appearance~\cite{Carena:2017qhd} and disappearance~\cite{Stenico:2018jpl} channels could be affected at SBN.

{\bf Resonant $\nu_\mu\to\nu_e$ oscillations.} The presence of a light scalar boson that couples only to neutrinos could induce a MSW effect sourced by the  cosmic neutrino background. This new matter potential depends on the neutrino energy as it has contributions from a $s$-channel scalar exchange. This model could explain the short-baseline anomalies if the local cosmic neutrino background is considerably larger than na\"ive expectations~\cite{Asaadi:2017bhx}. At SBN, $\nu_\mu\to\nu_e$ transitions would be present but the $\nu_\mu$ disappearance would be quite suppressed with respect to the usual 3+1 sterile neutrino scenario.

\begin{figure}[t]
\centering
\begin{minipage}{.32\textwidth}
  \centering
\includegraphics[width=0.95\textwidth]{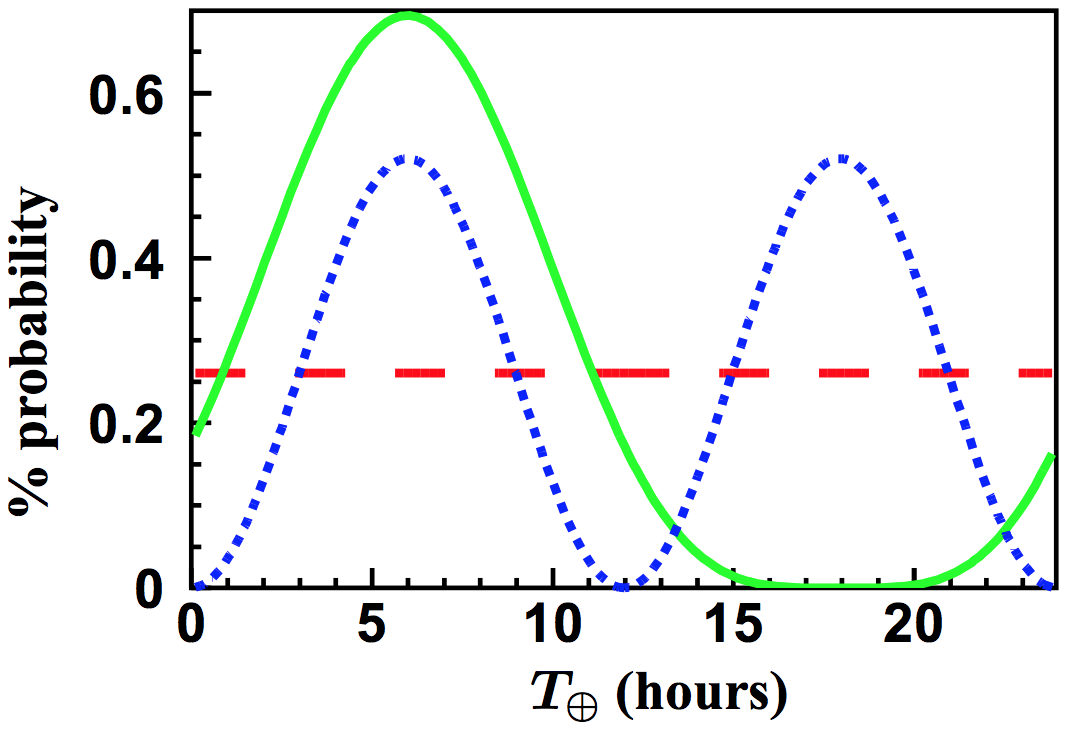}
\end{minipage}
\begin{minipage}{.32\textwidth}
  \centering
  \includegraphics[width=0.95\textwidth]{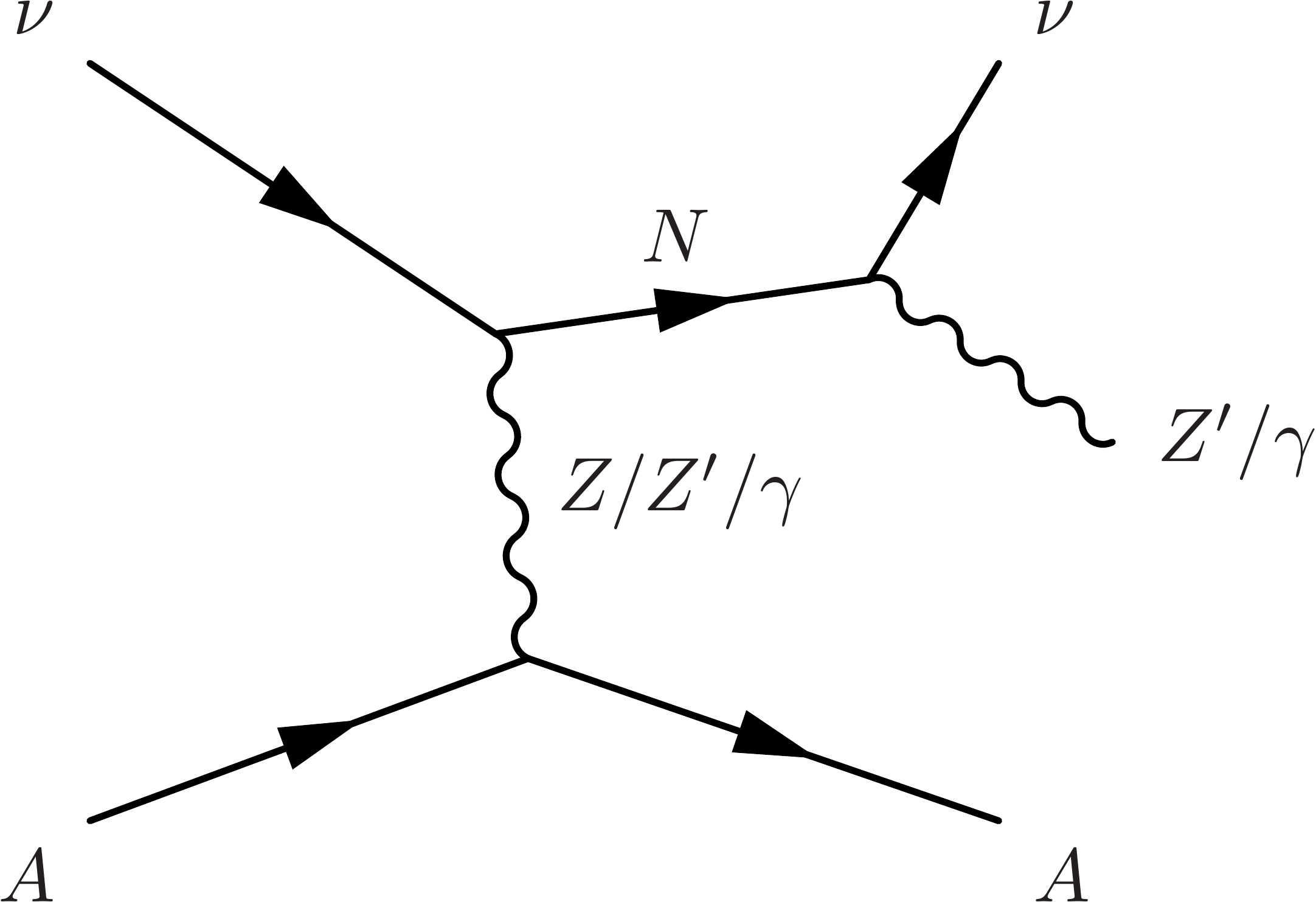}
\end{minipage}%
\begin{minipage}{.35\textwidth}
  \centering
\includegraphics[width=0.95\textwidth]{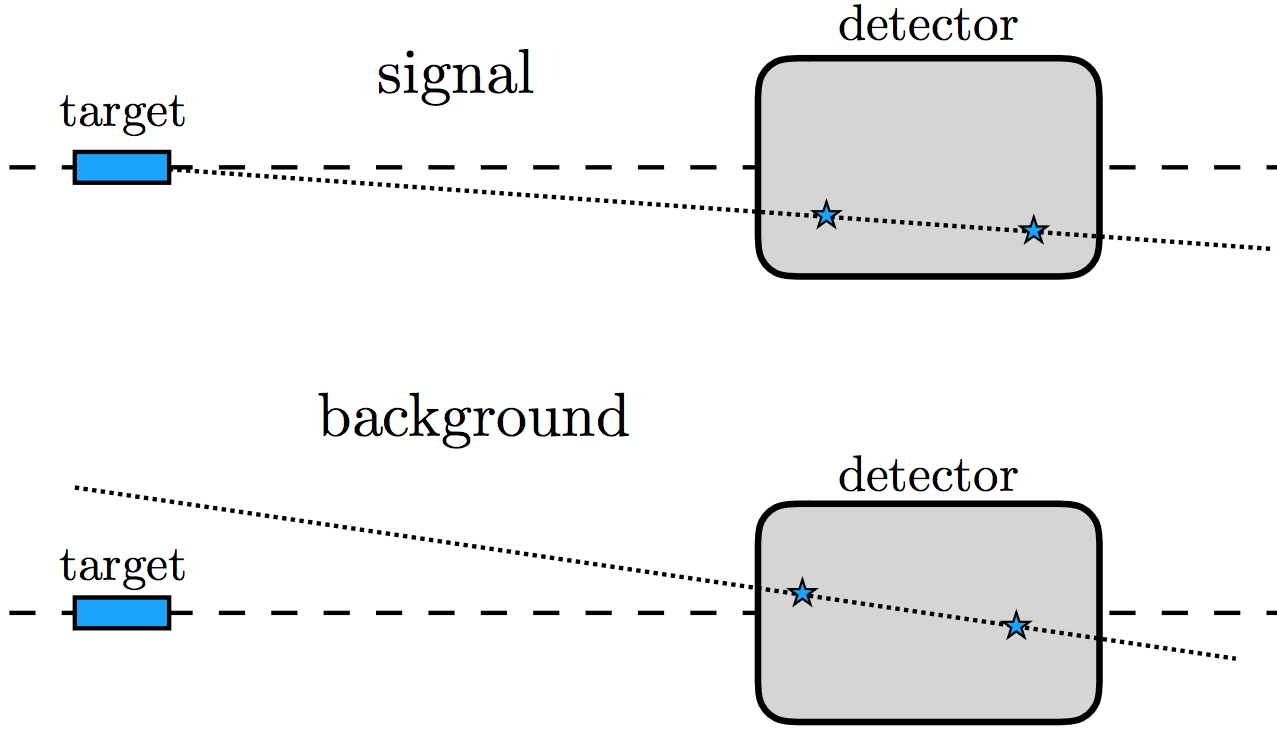}
\end{minipage}
\caption{{\it Left:}~Example of short-baseline $\bar\nu_\mu\to\bar\nu_e$ transition probability in a 3-neutrino model with Lorentz and CPT violation  as a function of sidereal time~\cite{Kostelecky:2004hg}.  The dashed red, solid green and dotted blue lines are three different realizations of the model where the average probability for the non-standard transition $\bar\nu_\mu\to\bar\nu_e$ at the LSND experiment is  $0.26\%$.
Figure taken from Ref.~\cite{Kostelecky:2004hg}. {\it Middle:}~The relevant process for dark neutrino sectors and transition magnetic moment scenarios. {\it Right:}~Low background signature of millicharged particles in SBN. Figure adapted from Ref.~\cite{Harnik:2019zee}.}
\label{fig:bsm}
\end{figure}

{\bf Violation of Lorentz and CPT symmetry} may be present in extensions of the Standard Model such as string theory~\cite{Kostelecky:1988zi}. The observable effects of these scenarios at SBN are modifications in the oscillation probability, such as direction dependent effects, neutrino-antineutrino mixing, annual modulations, and energy dependent effects on mass splittings and mixing angles (see e.g. Refs.~\cite{Kostelecky:2003cr, Kostelecky:2004hg, Katori:2006mz, Diaz:2011ia}). The left panel of Fig.~\ref{fig:bsm} shows an example of the phenomenology in short-baseline experiments.

{\bf Sterile neutrinos and altered dispersion relations} (ADR) were proposed as an explanation of the short-baseline anomaly~\cite{Pas:2005rb}. The idea is that additional terms are present in the dispersion relation for eV-scale sterile neutrinos $E^2 = |\vec{p}|^2+m^2$, like  the usual MSW effect, but coming from possible violations of the Lorentz symmetry or extra dimensional setups. In this scenario, the short-baseline anomalies would be explained by active-sterile neutrino oscillations. An energy dependent effective potential, generated by the ADR, weakens the constraints from some high energy  experiments like MINOS/MINOS+ and IceCube by reducing the active-sterile mixing. At SBN, this model would have a similar phenomenology to a usual 3+3 sterile scenario.

{\bf Heavy sterile neutrinos} are present in many extensions of the SM, possibly playing an important role in the  neutrino mass mechanism. Such states, if light enough, can be produced by meson decays via mixing with active neutrinos, propagate to and decay inside the detector (see, e.g. Ref.~\cite{Ballett:2016opr}). Typical decay signatures consist of $\pi^+ \ell^-$, $\nu\pi^0$, $\nu\ell^+\ell^-$, and $\nu\ell^+\ell^{\prime-}$. If the sterile neutrino is heavy enough, its signal will be slightly delayed with respect to the beam neutrino signal, which can be used to reduce backgrounds. Besides, the angular distribution of some decay products can reveal the Dirac/Majorana character of these heavy states~\cite{Balantekin:2018ukw}.

{\bf Charged current non-standard interactions} (CCNSI) in the lepton sector can be parametrized by effective operators such as $(\bar\ell_\alpha \Gamma \nu_\beta) (\bar f \Gamma f')/\Lambda^2$, where $\alpha,\,\beta$ are flavor indices, $f,\,f'$ denote SM fermions, $\Gamma$ is the Lorentz structure and $\Lambda$ is the scale of new physics. These operators can have many observable effects on neutrino detection, namely, (1) deviations of the SM CC quasi-elastic cross section, (2) modification of angular and energy distributions due to the presence of new Lorentz structures, and (3) flavor violation such as $\nu_\mu n \to e^- p$; see e.g. Refs.~\cite{Ohlsson:2012kf, Miranda:2015dra, Farzan:2017xzy} and references therein, as well as \cite{Liao:2018mbg} for a possible explanation of the short-baseline anomalies. At SBN, CCNSIs can induce apparent $\nu_\mu\to\nu_e$ conversion independently of the baseline.

{\bf Dark neutrino sectors.} A low-scale, dynamical mechanism of neutrino masses presents a deep connection between neutrinos and the dark sector~\cite{Bertuzzo:2018ftf}. In this class of models, right-handed neutrinos are charged under a new gauge symmetry, leading to neutrino upscattering into a heavy state which then decays to a light neutrino and a gauge boson, followed by the gauge boson decay to visible particles, that is, $\nu\,A\to N\,A\to \nu Z' A\to \nu\,\ell^+\ell^- A$, where $A$ denotes a nucleus, see middle panel of Fig.~\ref{fig:bsm}. Given an appropriate mass spectrum (with particles between the MeV and GeV scales), this model may yield the MiniBooNE low energy excess, presenting excellent agreement with angular and energy spectra~\cite{Bertuzzo:2018itn, Ballett:2018ynz}. At SBN, typical signatures of this framework would be pair production of $e^+e^-$, $\mu^+\mu^-$ or $\pi^+\pi^-$, induced by neutrino interactions, with little to no hadronic activity. The signal would be present at all three detectors, since there is no $L/E$ dependence.

{\bf Heavy neutrinos and transition magnetic moment.} A magnetic dipole moment $\mu_{\rm tr}$ induces light to heavy neutrino transitions via the effective operator $\mu_{\rm tr}\bar N_R \sigma_{\mu\nu} \nu_L F^{\mu\nu}$, where $F^{\mu\nu}$ is the electromagnetic field strength and $\sigma_{\mu\nu}\equiv\frac{i}{2}[\gamma_\mu,\gamma_\nu]$. This interaction was proposed as an explanation of the LSND/MiniBooNE excess~\cite{Gninenko:2010pr} where a muon neutrino would upscatter to $N$ followed by its decay to a photon, see middle panel of Fig.~\ref{fig:bsm} (see also Refs.~\cite{Masip:2012ke, Alvarez-Ruso:2017hdm}). The model signature at SBN would be  single photon production with small hadronic activity. LArTPC $e$-$\gamma$ discrimination capability  places SBN in a special position to probe these scenarios~\cite{Magill:2018jla}.  The signal would be present at all three detectors, since there is no $L/E$ dependence.

{\bf Neutrino tridents} are Standard Model processes in which a neutrino produces a pair of leptons, $\nu\,A\to\nu\,A \,\ell^+\ell'^-$, although only the $\mu^+\mu^-$ trident was observed so far. These processes are very sensitive to the presence of new physics, especially light states below the weak scale, and thus provide an invaluable tool to search for new physics~\cite{Altmannshofer:2014pba, Magill:2017mps, Ballett:2018uuc}. The signature at SBN would be a pair of leptons, of same or different flavor, induced by neutrino interactions, and with little or no hadronic activity. Trident events from light dark sector particles would lead to similar same-flavor signatures at SBN~\cite{deGouvea:2018cfv}.

{\bf Millicharged particles}~\cite{Dobroliubov:1989mr} with fractional electric charge can arrive in simple extensions of the Standard Model. Millicharged particles (mCPs) can be produced via electromagnetic interactions in meson decays and Drell-Yan processes. At SBN, two promising signatures have been identified. Elastic scattering of mCPs with electrons~\cite{Magill:2018tbb}, as it proceeds via photon exchange, is enhanced at low electron recoils which can be uniquely probed in LArTPCs~\cite{Acciarri:2018myr} (although neutron induced backgrounds are also large). A cleaner way to probe mCPs with much lower backgrounds is by looking at multiple low energy deposition hit events aligned with the beam target~\cite{Harnik:2019zee}, see right panel of Fig.~\ref{fig:bsm}. The multiple hit requirement greatly reduces the backgrounds, but also reduces signal. Since the neutrino detectors have different size and cosmic ray background, signal-to-noise optimization is needed for each of them.

{\bf Light dark matter} coupling to a new, light gauge boson that mixes with the photon can be produced in neutrino beams by meson decays and then propagate to and interact with neutrino detectors (see e.g. Refs.~\cite{Batell:2014yra, Coloma:2015pih, Frugiuele:2017zvx, Kim:2018veo}). The dark matter signal in this scenario is very similar to neutrino NC scattering, and thus neutrino events are the main background. Triggering on highly off-axis beams may help reduce backgrounds, as well as analyzing spectral information or running the experiment in beam dump mode (see the recent analysis performed by the MiniBooNE collaboration~\cite{Aguilar-Arevalo:2018wea}). Variations of this model where the dark particle decays to a dark photon and a lighter dark matter species can be probed in SBN by looking at $e^+e^-$ pairs with no hadronic activity~\cite{Jordan:2018gcd}. 

{\bf Neutrinophilic, lepton-number charged scalars} may exist as a portal to dark matter. Such a scalar $\phi$ couples to neutrinos via the effective operator $(LH)(LH)\phi/\Lambda^2$, possibly leading to the scattering process $\nu_\mu + p \to \ell^+ \phi^* n$, which can be measured especially well in a LArTPC~\cite{Berryman:2018ogk}. This process has two characteristics of specific interest to the SBN program. First, the event, due to the $\phi^*$ in the final state, appears to have a large amount of missing transverse momentum, despite appearing like a normal charged-current process otherwise.  Second, the final state has a ``wrong-sign'' lepton, which may be possibly exploited  by charge identification on a statistical basis.  Precise reconstruction of the final state in a LArTPC can provide a way to search for this process as well.

\vspace{0.5em}

The SBN program has the potential to probe a vast range of beyond the Standard Model physics: sterile neutrinos from the eV to the MeV scales, new interactions, extra dimensions, violations of Lorentz and CPT symmetries, dark matter, and so on.  We can thus appreciate that the excellent LArTPC event reconstruction and large statistics lead to a rich physics program at SBN.

%%%%%%%%%%%%%%%%%%%%%%%%%%%%%%%%%%%%%%%%%%%%%%%%%%%%%%%%%%%%%%%%%%%%%%%%%%%%%%
\section{SUMMARY AND OUTLOOK}
\label{sec:conclusion}
%%%%%%%%%%%%%%%%%%%%%%%%%%%%%%%%%%%%%%%%%%%%%%%%%%%%%%%%%%%%%%%%%%%%%%%%%%%%%%

In the next few years, the SBN program at Fermilab will provide powerful new input to the question of light sterile neutrinos.  Three large precision detectors with the identical nuclear target and same detection technique all sitting in a single neutrino beam allows for a level of control of systematic uncertainties that will be unprecedented in sterile neutrino searches. Furthermore, the experiment is simultaneously sensitive to both $\nu_e$ appearance and $\nu_\mu$ disappearance oscillation channels, another first and a key ingredient to solving the light sterile neutrino puzzle.  
Not limited to the sterile neutrino question, SBN has a broad range of physics goals that include detailed, high-statistics studies of neutrino-nucleus scattering (argon specifically, with significant importance for the DUNE experiment in the future) and the exploration of a range of exciting beyond the Standard Model theories that can potentially be tested with SBN data. 
Finally, SBN is a valuable opportunity for the large international community developing the challenging techniques needed to extract physics information from \lartpc data and that is now making plans to construct and operate these detectors at enormous scales. 

Preparation of the SBND and ICARUS detectors is well underway at the time of publication of this article.  ICARUS was installed in the experimental hall along the beam in August 2018 and is being readied for liquid argon fill by late in 2019.  SBND construction is continuing apace with all major detector components currently completed or in production.  The TPC components have all arrived to Fermilab where detector assembly, which is done outside of the cryostat, will be completed by fall 2019 and the detector made ready for transport to the near detector hall along the beam.  \sbnd will be ready for argon fill by the end of 2020, officially beginning the era of full SBN program operations.       

\begin{summary}[SUMMARY POINTS]
\begin{enumerate}
\item The SBN program at Fermilab will address the `short-baseline anomalies', which could be hinting at the possible existence of new sterile neutrino states.
\item SBN uses multiple large liquid argon TPC neutrino detectors, the key to its world-leading sensitivity to oscillations.
\item The SBN detectors are being realized by a large international team of scientists and represent a valuable development platform for this technology, in preparation for the DUNE/LBNF long-baseline program. 
\item In addition to sterile neutrinos, the detectors and beams of the SBN program enable a broad science program of neutrino-argon interactions and searches for physics beyond the Standard Model.
\end{enumerate}
\end{summary}

\begin{issues}[OUTLOOK]
\begin{enumerate}
\item The SBN far and near detectors will begin full physics operation by early 2020 and 2021, respectively, leading immediately to non-oscillation physics and a ruling on the light sterile neutrino question within a few years.  
\end{enumerate}
\end{issues}

% Acknowledgements
\section*{ACKNOWLEDGMENTS}
We thank Roni Harnik, Kevin Kelly and Danny Marfatia for useful discussions, and Mona Dentler and Rhiannon Jones for their input and help with figures. We also acknowledge the contributions of all the SBN collaborators to the physics studies and detectors discussed here. PM thanks the organizers of the DCPIHEP workshop 2019 and the warm hospitality of the Universidade de S\~{a}o Paulo where a large fraction of this work was completed.  Fermilab is operated by the Fermi Research Alliance, LLC under contract No. DE-AC02-07CH11359 with the United States Department of Energy. 

% References
\bibliographystyle{ar-style5}
\bibliography{refs}

\end{document}